\documentclass[useAMS,usenatbib]{mn2e}
\usepackage{graphicx}

\newcommand{\bm}[1]{\mbox{\boldmath{$#1$}}}

\title[Rotating axisymmetric sunspots]
      {Numerical simulations of rotating axisymmetric sunspots}
\author[Botha, Busse, Hurlburt \& Rucklidge]
      {G. J. J. Botha$^{1}$\thanks{E-mail: gert@maths.leeds.ac.uk}, 
       F. H. Busse$^{2}$\thanks{E-mail: fbusse@igpp.ucla.edu}, 
       N. E. Hurlburt$^{3}$\thanks{E-mail: hurlburt@lmsal.com} and 
       A. M. Rucklidge$^{1}$\thanks{E-mail: A.M.Rucklidge@leeds.ac.uk}\\
$^{1}$Department of Applied Mathematics, University of Leeds, 
      Leeds, LS2 9JT, UK\\
$^{2}$Institute of Geophys. Planet. Physics, UCLA, Los Angeles, 
      CA 90024, USA\\
$^{3}$Lockheed Martin Solar and Astrophysics Laboratory, 
      Organization ADBS Building 252, Palo Alto, CA 94304, USA}
\begin{document}

\date{Accepted 12 April 2008; First submitted 22 February 2008}

\pagerange{\pageref{firstpage}--\pageref{lastpage}} \pubyear{2007}

\maketitle

\label{firstpage}

\begin{abstract}
A numerical model of axisymmetric convection in the presence of a
vertical magnetic flux bundle and rotation about the axis is presented. 
The model contains a compressible plasma described 
by the nonlinear MHD equations, with density and temperature gradients 
simulating the upper layer of the sun's convection zone. 
The solutions exhibit a central magnetic flux tube in a cylindrical numerical 
domain, with convection cells forming collar flows around the tube.  
When the numerical domain is rotated with a constant angular velocity, 
the plasma forms a Rankine vortex, with the plasma rotating 
as a rigid body where the magnetic field is strong, 
as in the flux tube, while experiencing sheared azimuthal flow 
in the surrounding convection cells, forming a free vortex. As a result, 
the azimuthal velocity component has its maximum value close to the outer 
edge of the flux tube.
The azimuthal flow inside the magnetic flux tube and the vortex flow are  
prograde relative to the rotating cylindrical reference frame. A retrograde 
flow appears at the outer wall. 
The most significant convection cell outside the flux tube is the location 
for the maximum value of the azimuthal magnetic field component. 
The azimuthal flow and magnetic structure are not generated spontaneously, 
but decay exponentially in the absence of any imposed rotation of the 
cylindrical domain.
\end{abstract}



\section{Introduction}

Observations of sunspots rotating around their own axis, which is 
perpendicular to the plane of the photosphere,
have a long history throughout the twentieth century. 
\citet{Hale08} and \citet{Evershed10} noticed the rotation as well as 
a vortex forming around the rotating sunspot. Observations in the  
photosphere and corona continued through the century, culminating in 
the high resolution measurements of today. (See \citet{BrownEA03} 
and references therein.) An exciting development during the last decade 
has been the ability to measure the associated flow beneath 
the photosphere \citep{GizonBirch05,ZhaoKosovichev03}. 

There is a distinct radial profile associated with the azimuthal velocity 
of rotating sunspots. \citet{BrownEA03} found that the 
umbra (in which the rotation axis resides) has small average azimuthal 
velocities, while the fastest rotation occurs at some point along the radial 
length of the penumbra. The rotation then tails away to a negligible value 
outside of the sunspot. The peak azimuthal velocity in the penumbra can be 
more than double that inside the umbra. \citet{BrownEA03} found suggestions 
of rotation outside some sunspots, but these observations are hampered by an 
ambiguous penumbral edge. 
In contrast, \citet{YanQu07} observed a rotating sunspot where the  
maximum azimuthal velocity occurred inside the umbra. The rotation persisted 
in the penumbra and the area near the penumbra, with the angular velocity 
reducing as one moves radially away from the umbra. The surrounding area 
far removed from the penumbra experienced a slow rotation in the opposite 
direction from that of the rotating sunspot. 

It is not clear if the direction of rotation has a hemispheric preference. 
\citet{Knoska75}, and references therein, found that the majority of 
rotating sunspots in both hemispheres turn anticlockwise. However,  
\citet{DingEA87} found a preference, with clockwise (anticlockwise) 
rotation predominantly in the southern (northern) hemisphere. 
This would suggest that the Sun's differential rotation associated with 
the global flow field has an influence. The Coriolis force due to the 
Evershed flow field would cause rotation in the opposite direction. 
However, helioseismic observations \citep{GizonBirch05}, supported by 
numerical results \citep{HurlburtRucklidge00,BothaEA06}, 
show a converging horizontal flow below the Evershed flow, which leads to 
cyclonic vorticity and hence a possible contribution from the Coriolis force. 

The small sample of rotating sunspots studied by \citet{BrownEA03} 
suggests that younger sunspots rotate faster than older ones. However, 
it was difficult to judge the ages of the sunspots in the sample. 
The rotation rates are time dependent, with all rotation eventually 
decreasing with time. The peak rotation in the penumbra can be anything 
up to 3 degrees per hour, as observed by \citet{BrownEA03}. 
The same time evolution was observed in a 
rotating pore \citep{DorotovicEA02}. This behaviour suggest that some of the 
rotation are caused by local events that are transitory in nature. This 
conclusion is strengthened by an observation of damped oscillatory motion, 
which had a maximum rotation of 3.5 degrees per hour \citep{Kucera82}. 

A possible mechanism causing rotating sunspots is the rise of twisted flux 
ropes \citep{GibsonEA04}. In this model the rotation of two flux rope poles is 
observed after the central horizontal portion of the flux rope has emerged 
through the photosphere. This implies the existence of two co-evolving 
sunspots of opposite magnetic polarity. This is generally not in evidence 
in the observations, mostly because leading sunspots are often followed by a 
more diffuse opposite polarity.  

It was shown by \citet{GupasyukGopasyuk05} that when the averaged velocity 
and magnetic field components are subtracted from observed sunspots, 
the result fits a lightly damped sinusoidal wave. This implies that as well 
as the motion described so far, sunspots also experience torsional 
oscillations. Using the thin flux tube model, \citet{MusielakEA07} 
found that in a compressible, isothermal field-free medium, linear 
torsional Alfv\'en waves along the magnetic tube do not have a cutoff 
frequency. 

The rotation of sunspots has been linked to the formation of soft X-ray 
sigmoids as well as the eruption of flares 
\citep{Alexander06,RegnierCanfield06,TianAlexander06}. 
Numerical simulations by \citet{GerrardEA03} 
show that by adding a horizontal photospheric flow to the rotation, 
the generation of flares is  
enhanced as both rotation and flow increase the complexity of the magnetic 
field. This is supported by the observation that flare activity is 
correlated to magnetic flux and kinetic vorticity \citep{MasonEA06}. 

Evidence from helioseismic measurements show that the 
rotation of sunspots, as observed in the photosphere, extends into the 
deeper layers of the sun. Up to a depth of approximately 7 Mm vortical flow 
in the same direction as the rotation of the sunspot exists, while below 7 Mm 
a vortical flow opposite to the sunspot rotation direction is observed 
\citep{Kosovichev02,ZhaoKosovichev03,GizonBirch05}. 
However, it should be noted that helioseismic measurements are difficult 
and not always consistent. For example, up to a depth of 3 Mm a 
converging inflow is found when using p modes, while f mode measurements 
find only outflows down to 10 Mm \citep{GizonBirch05}.

In this paper an axisymmetric model is used to simulate rotation 
around a central magnetic flux bundle. The values of the physical 
parameters of the model are chosen to describe the solar convection zone 
from a depth of approximately 500 km below the visible surface of the Sun 
to a depth of approximately 6000 km \citep{BothaEA06}. The numerical 
domain is a cylinder with an aspect ratio of $\Gamma=3$, i.e.\ one unit 
deep and a radial distance of three units. This implies that we are 
simulating magnetoconvection on the supergranular scale. 
To generate azimuthal flow and magnetic field, the whole domain is 
rotated at a constant angular velocity. 
Strictly speaking this is not equivalent to simulating a pore or 
sunspot where only the magnetic flux bundle rotates. However, in spite of 
driving the azimuthal flow throughout the numerical domain, we find 
that the solution tends to conform to observations of \citet{BrownEA03},  
where a maximum azimuthal flow occurs close to the magnetic 
flux bundle. This means that a vortical flow forms around the 
flux bundle while the plasma inside the bundle rotates as a solid body. 
This type of flow is formally described as a Rankine vortex.  

Our numerical results may be compared to results found by 
\citet{JonesGalloway93}, who studied a Boussinesq fluid in an 
axisymmetric cylinder.
They imposed two types of boundary conditions: a stress-free outer wall 
as well as an external flow, implemented by rotating the outer wall of the 
fixed cylinder. A flux bundle formed at the central axis, with the 
maximum angular velocity occurring near the axis. For high magnetic 
field strength, described by the dimensionless Chandrasekhar number ($Q$), 
the flux bundle broadened with the stress-free boundary conditions  
producing a reverse in azimuthal magnetic flux near the axis, while the 
imposed external flow produced a reverse azimuthal flow near the axis. 
The reversal in the azimuthal magnetic field is ascribed 
to the conservation of angular momentum under stress-free conditions, 
while the flow reversal obtained with the imposed external flow is 
ascribed to the working of the Lorentz force. 

In Section \ref{sec:model} the mathematical model and its numerical 
implementation is described. This is followed by the numerical results. 
When no rotation of the numerical domain is present (Section 
\ref{sec:norot}), no azimuthal velocity and magnetic field are generated 
spontaneously: both quantities are small and decay exponentially with time. 
This case is useful to compare the rotating solutions against. 
Driven by the rotating numerical domain, the solution settles into a time 
independent solution that shows rigid rotation of the plasma in the magnetic 
flux tube and vortical rotation around it (Sections \ref{sec:rot1} and 
\ref{sec:rot2}). 
This solution is robust and essentially stays the 
same when the magnetic field strength is increased (Section 
\ref{sec:mag}), the Prandtl number is lowered (Section \ref{sec:Prandtl}), 
as well as when the stratification is increased (Section \ref{sec:depth}).
The latter part of the numerical investigation explores the influence 
of the numerical domain on the solution. This we do by changing the 
bottom and outside boundary conditions (Sections \ref{sec:Tflux} and 
\ref{sec:out}). We conclude the paper with a summary of the results. 


\section{Model}
\label{sec:model}

Partial differential equations describing compressible magnetoconvection 
are solved in an axisymmetric cylindrical geometry, using a numerical code 
developed for this purpose. A detailed description of the two dimensional 
(2D) model is given by \citet{HurlburtRucklidge00}. 
Here we extend the model to 2.5D by including azimuthal components in 
addition to the radial and axial components. A constant angular velocity is 
added that introduces the Coriolis and centrifugal forces into the 
Navier-Stokes equation. 

\subsection{Mathematical model}

The initial temperature and density profiles in the vertical ($z$) 
direction are given by  
\begin{eqnarray}
T&=&1+\theta z, \label{eq:T} \\
\rho&=&(1+\theta z)^m . \label{eq:rho}
\end{eqnarray}
The temperature and density are scaled so that they are equal to 1 at the 
top of the static atmosphere. The initial temperature gradient is given 
by $\theta$, while $m$ is the polytropic index. 
The equations for fully compressible, nonlinear axisymmetric 
magnetoconvection that we use are 
\begin{eqnarray}
\frac{\partial\rho}{\partial t}
&=&
-\nabla\cdot({\bf u}\rho) ;
\\
\frac{\partial {\bf u}}{\partial t}
&=& 
-{\bf u}\cdot\nabla{\bf u}
-2{\bf\Omega}\times{\bf u} 
+ \Omega^2(\hat{\bf z}\times{\bf r})\times\hat{\bf z}
-\frac{1}{\rho}\nabla P
\nonumber \\ & & 
+\theta(m+1)\hat{\bf z}
+\frac{\sigma K}{\rho}\nabla\cdot\tau
-\frac{\sigma\zeta_0 K^2Q}{\rho}{\bf j}\times{\bf B} ;
\label{eq:NS}
\\
\frac{\partial T}{\partial t}
&=&
-{\bf u}\cdot\nabla T
-(\gamma-1)T\nabla\cdot{\bf u}
+\frac{\gamma K}{\rho}\nabla^2T
\nonumber \\ & & 
+\frac{\sigma K(\gamma -1)}{\rho}
\left(\frac{1}{2}{\bf\tau}:{\bf\tau}+\zeta_0^2QK^2j^2\right) ;
\\
\frac{\partial A_\phi}{\partial t}
&=&
({\bf u}\times{\bf B})_\phi
-\zeta_0 Kj_\phi ;
\\
\frac{\partial B_\phi}{\partial t}
&=& 
\Big[ \nabla\times({\bf u}\times{\bf B})\Big]_\phi 
+\zeta_0 K\left( \nabla^2 B_\phi-\frac{B_\phi}{r^2}\right) .
\label{eq:Bphi}
\end{eqnarray}
The cylindrical reference frame is rotated about its axis at a constant 
angular velocity of ${\bf\Omega}=(d\phi/dt)\hat{\bf z}$, which is 
responsible for the Coriolis and centrifugal terms in the Navier-Stokes 
equation (\ref{eq:NS}). 
The vector potential $A_\phi$ gives the $r$ and $z$ components of the 
magnetic field while the azimuthal component is included explicitly, 
so that the magnetic field is given by 
\begin{equation}
{\bf B}=\nabla\times(\hat{\bm\phi}A_\phi) 
         + \hat{\bm\phi}B_\phi.
\end{equation} 
The velocity consists of three components, namely ${\bf u}=(u_r,u_\phi,u_z)$, 
where $u_\phi$ refers to the azimuthal velocity relative to the rotating 
reference frame.  
We also use the auxiliary equations
\begin{equation}
\nabla\cdot{\bf B}=0, \hspace{1cm}
P=\rho T,\hspace{1cm}
{\bf j}=\nabla\times{\bf B} , 
\label{eq:aux}
\end{equation}
and the following notation:  
$\gamma$ is the ratio of specific heats; 
$\sigma$ the Prandtl number;  
$\zeta_0$ the magnetic diffusivity ratio at $z=0$; 
and the Chandrasekhar number given by 
\begin{equation}
Q=\frac{(Bd)^2}{\mu\rho\eta\nu}\, , 
\end{equation}
where $d$ is the depth of the domain, $\mu$ the magnetic permeability, 
$\eta$ the magnetic diffusivity and $\nu$ the kinetic viscosity. 
The rate of strain tensor is given by 
\begin{equation}
{\bf\tau}=
\left[\begin{array}{ccccc}
      \displaystyle{ 2\frac{\partial u_r}{\partial r} }
  &  & \displaystyle{ \frac{\partial u_\phi}{\partial r}-\frac{u_\phi}{r} }
  &  & \displaystyle{ \frac{\partial u_r}{\partial z}+\frac{\partial u_z}{\partial r} }
   \\
  & & & & \\
      \displaystyle{ \frac{\partial u_\phi}{\partial r}-\frac{u_\phi}{r} }
  &  & \displaystyle{ 2\frac{u_r}{r} }
  &  & \displaystyle{ \frac{\partial u_\phi}{\partial z} }
   \\
  & & & & \\
      \displaystyle{ \frac{\partial u_r}{\partial z}+\frac{\partial u_z}{\partial r} }
  &  & \displaystyle{ \frac{\partial u_\phi}{\partial z} }
  &  & \displaystyle{ 2\frac{\partial u_z}{\partial z} }
      \end{array}\right] , 
\label{eq:strain}
\end{equation}
while the dimensionless thermal conductivity $K$ is related to the Rayleigh 
number $R$ in the following way: 
\begin{equation}
R=\theta^2(m+1)\left[ 1-\frac{(m+1)(\gamma-1)}{\gamma}\right]
  \frac{(1+\theta/2)^{2m-1}}{\sigma K^2} \; .
\label{eq:R}
\end{equation}
$R$ is a measure of the importance of buoyancy forces compared to viscous 
forces in the middle of the layer, and is used to drive the convection in the 
model. The addition of rotation adds an additional scaling which relates the 
convective timescale to the rotational timescale. Following \citet{Gilman77} 
we express this ratio using the convective Rossby number defined as
\begin{equation}
Ro={\sigma K  \over 2 \Omega}\sqrt{R}.
\label{eq:Ro}
\end{equation}
\citet{BrummellEA96} found that this parameter, which can be 
evaluated from the control parameters, is typically close to that based on 
the traditional definition of the Rossby number, namely the ratio of 
the rms vorticity of the flow to the vorticity $2\Omega$ associated
with the rotating cylinder. The effects of rotation are significant for 
$Ro \approx 1$, and dominate for $Ro\ll1$. For supergranules and sunspot 
moats, $Ro \approx 30$.

All the other symbols have their usual meaning. 
The physical quantities are dimensionless, with the length scaled 
proportional to the depth of the numerical domain, velocity scaled 
proportional to the sound speed at the top of the domain 
and temperature, magnetic field, density, and pressure all scaled 
proportional to their initial values at the top of the numerical domain.
These top initial values are radially uniform and do not change throughout 
this paper, as discussed in Section \ref{sec:num}. 


\subsection{Numerical implementation}
\label{sec:num}

The computational domain is an axisymmetric cylinder of radius $\Gamma$, 
situated in the $(r,z)$ plane so that 
\begin{equation}
0\leq r\leq \Gamma, \hspace{1cm}
0\leq z\leq 1,
\end{equation}
with $z=0$ at the top of the box
\citep{HurlburtRucklidge00,BothaEA06}. 
We require that all variables be sufficiently well-behaved at the 
axis ($r=0$) and that the differential operators in the PDEs are 
non-singular. This implies that 
\begin{eqnarray}
\frac{\partial\rho}{\partial r}
=u_r=u_\phi=\frac{\partial u_z}{\partial r}
=\frac{\partial T}{\partial r}
=0 , & & \nonumber \\
A_\phi
=B_r=B_\phi=\frac{\partial B_z}{\partial r}
=j_\phi
=0 .& & 
\label{eq:axis}
\end{eqnarray}
Terms like $u_r/r$, $u_\phi/r$ and $B_\phi/r$ are evaluated using 
l'H\^opital's rule, while terms containing $u_r/r^2$ cancel algebraically. 

The outside wall ($r=\Gamma$) is a slippery, impenetrable  
wall with no lateral heat flux across it (i.e.\ an insulator): 
\begin{equation}
\frac{\partial T}{\partial r}=u_r=\frac{\partial u_z}{\partial r}=
B_r=B_\phi=j_\phi=0, 
\hspace{0.5cm}
\frac{\partial u_\phi}{\partial r}=\frac{u_\phi}{r}. 
\label{eq:outer}
\end{equation}
The magnetic potential has the value $A_\phi=\Gamma/2$ at the outside wall, 
which was chosen so that the initial vertical uniform field satisfies $B_z=1$. 

At the bottom boundary the magnetic field is vertical. The temperature 
$T$ is chosen to be constant with value $\theta+1$ from equation 
(\ref{eq:T}). The bottom boundary is impenetrable and stress free, i.e. 
\begin{equation}
B_r
=B_\phi
=\frac{\partial B_z}{\partial z}
=\frac{\partial u_r}{\partial z}
=\frac{\partial u_\phi}{\partial z}
=u_z=0. 
\label{eq:bottom}
\end{equation}

The top of the box is treated as impenetrable for the plasma, with a radiative 
temperature boundary condition given by Stefan's law: 
\begin{equation}
\frac{\partial u_r}{\partial z}
=\frac{\partial u_\phi}{\partial z}
=u_z=0, 
\hspace{1cm}
\frac{\partial T}{\partial z}=\theta T^4. 
\label{eq:top}
\end{equation} 
The $\theta$ in Stefan's law is the same as in (\ref{eq:T}), so that 
the equilibrium profile used as initial condition is not destroyed. The 
Stefan-Boltzmann constant will enter (\ref{eq:top}) only when we 
dimensionalize it.  
The magnetic field is matched to a potential field on top of the numerical 
domain, $\partial A_\phi/\partial z=M_{pot}(A_\phi)$, where $M_{pot}$ is 
a linear operator, so that $B_r$ and $B_z$ are continuous across the boundary. 
The potential field is solved by assuming an infinitely tall conducting 
cylinder of radius $\Gamma$ above the domain, with the magnetic field becoming 
uniform as $z\rightarrow\infty$. A more detailed description is given by 
\citet{HurlburtRucklidge00}. 
No currents exist inside the potential field and consequently we choose 
$j_z=0$ along the top of the box. From (\ref{eq:axis}) it then follows that 
$B_\phi=0$ along the top boundary. 

One consequence of these boundary conditions is that no current escapes 
from the numerical domain: $j_r=0$ on $r=\Gamma$ and $j_z=0$ on $z=0$ and 1. 
It follows that the boundaries do not provide any net vertical torques, 
and so do not contribute to changes in the vertical component of the 
total angular momentum, $L_z$. Nonetheless, $L_z$ is not conserved in 
compressible convection: the Coriolis term can lead to changes in 
$L_z$, for example, when mass is transported to larger distances from 
the axis. (Note that this does not occur in incompressible convection.) 
A consequence of this is that as the solution evolves through time, $L_z$ 
tends to drift, making the meaning of the parameter $\Omega$ less 
precise. Therefore, we will look for steady solutions with no net 
vertical angular momentum relative to the rotating frame. We achieve this 
by calculating the drift in the value of $L_z$ after each iteration and 
then introducing a correction in the form of an equivalent rigid body 
rotation in the opposite direction. This alters the evolution of the 
PDEs (slightly), but steady states are correct solutions of the PDEs. 
In our oscillatory solutions we remove the constraint from $L_z$ 
and follow its time evolution. This is discussed in Section \ref{sec:time}. 

A uniform, vertical magnetic field is used as initial condition. 
For a nonrotating cylinder ($\Omega=0$) the azimuthal magnetic field is 
perturbed (Section \ref{sec:norot}). For a finite angular velocity 
($\Omega\neq 0$) the evolution of the plasma is triggered by starting the 
quiet, nonrotating plasma with a finite $\Omega$ and no plasma perturbation. 
Both these initialisations ensure that $L_z=0$ at the start of the numerical 
simulations. 


\begin{figure}
\centerline{
\setlength{\unitlength}{0.6cm}
\begin{picture}(12,5)
\put(0,4){\line(1,0){6}}
\put(0,5){\line(1,0){6}}
\put(0.5,4.25){1. Lines}
\put(0,2){\line(1,0){6}}
\put(0.5,3.35){2. Lines}
\put(0.5,2.8){3. Arrows}
\put(0.5,2.25){4. Colour}
\put(0,0){\line(0,1){5}}
\put(0,0){\line(1,0){6}}
\put(6,0){\line(0,1){5}}
\put(0.5,1.35){5. Lines}
\put(0.5,0.8){6. Arrows}
\put(0.5,0.25){7. Colour}
\put(6.5,2){\line(1,0){6}}
\put(6.5,4){\line(1,0){6}}
\put(7.3,2.8){8. Colour}
\put(6.5,0){\line(0,1){4}}
\put(6.5,0){\line(1,0){6}}
\put(12.5,0){\line(0,1){4}}
\put(7.3,1.15){9. Colour}
\put(7.0,0.6){10. Arrows}
\end{picture}
 }
\caption{The diagnostics used to describe the numerical results: 
         1. Potential magnetic field lines; 
         2. Magnetic field lines; 
         3. Velocity field in the $(r,z)$ plane; 
         4. Temperature fluctuation relative to the unperturbed state; 
         5. Density contour lines; 
         6. Magnetic field direction and strength; 
         7. Azimuthal current density; 
         8. Azimuthal velocity field; 
         9. Azimuthal magnetic field; 
         10. Current density in the $(r,z)$ plane. 
         All colour scales have red as maximum and blue as minimum. 
         The colour green represents zero. 
         }
\label{fig:diag}
\end{figure}


\begin{figure}
\centerline{
\scalebox{0.4}
{\includegraphics{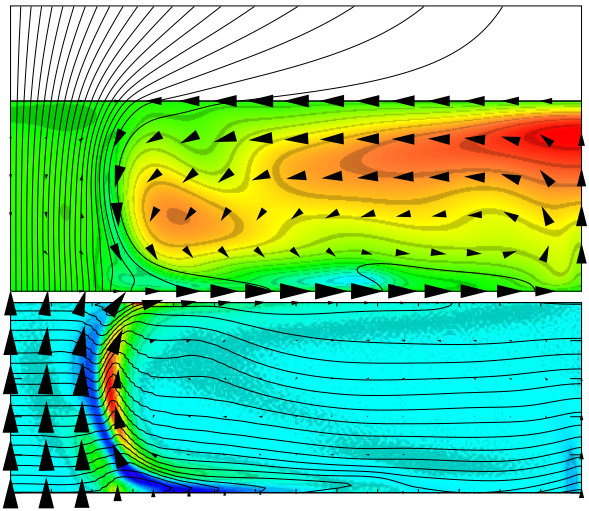}}
 }
\caption{No rotation with the azimuthal velocity and magnetic field 
         components approaching zero asymptotically. Consequently the 
         right hand side of Figure \ref{fig:diag} is not included. 
         The absolute temperature variation has a maximum of 
         $\max|\tilde{T}|=3.1$, while the azimuthal current density  
         $j_\phi$ lies in the interval (-200,350). 
         }
\label{fig:Q32om0}
\end{figure}


The density does not in principle satisfy a boundary condition, but we 
impose the value of the normal derivative of $\rho$ obtained from the 
Navier-Stokes equation (\ref{eq:NS}). 

The numerical code was developed specifically for this type of 
calculation \citep{HurlburtRucklidge00}. Sixth-order compact finite 
differencing is used, which reduces at the boundaries to fifth-order accuracy 
for first-order derivatives and fourth-order accuracy for second-order 
derivatives. 
The grid intervals were chosen to be equal in the $r$ and $z$ directions, 
with 240 grid points in the horizontal and 80 in the vertical for the 
majority of calculations. 
The time evolution obtained fourth-order accuracy through a modified 
(explicit) Bulirsch-Stoer integration scheme, with the time step limited by 
the Courant condition (using the maximum sound and Alfv\'en speeds, as well 
as the thermal diffusive limit), multiplied by a safety factor of 0.5.


\begin{figure*}
\centerline{
\begin{minipage}{17cm}
\begin{minipage}{8cm}
\scalebox{0.4}
{\includegraphics{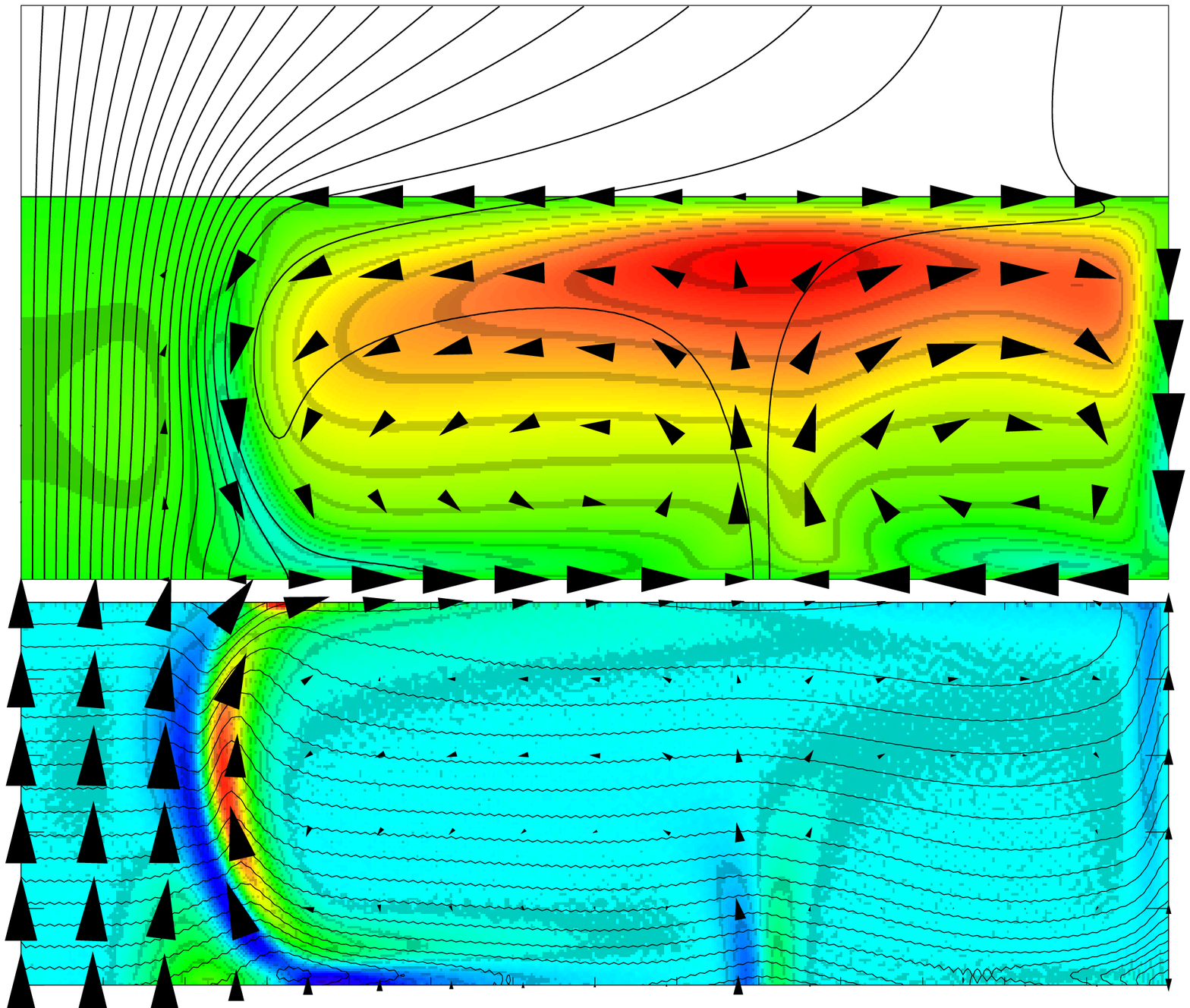}}
\end{minipage}
\begin{minipage}{8cm}
\scalebox{0.4}
{\includegraphics{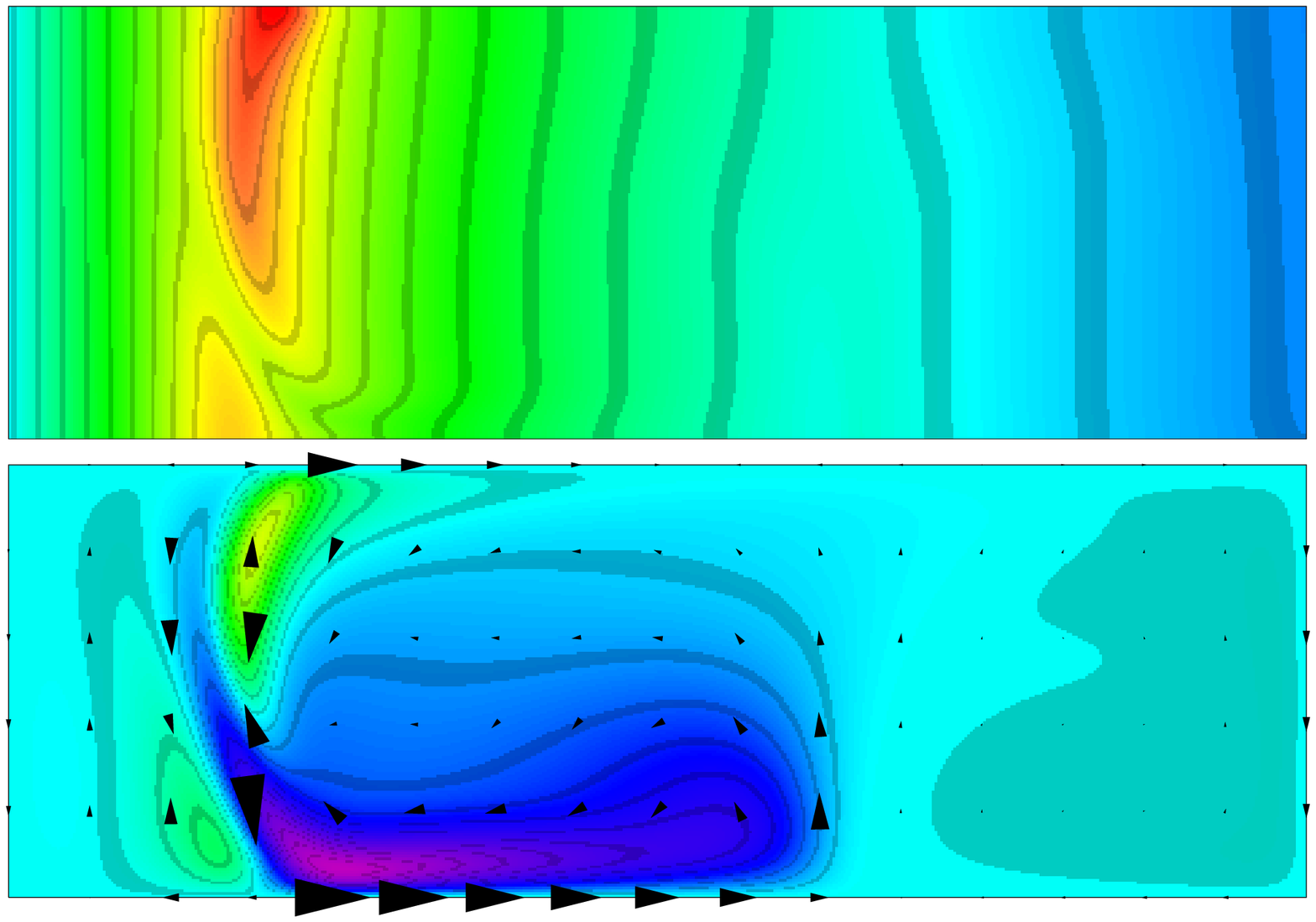}}
\end{minipage}
\end{minipage}
}
\caption{Constant angular velocity $\Omega=0.1$. 
         The time independent solution shows two convection cells in the 
         radial direction instead of the one in the result obtained with 
         no rotation (Figure \ref{fig:Q32om0}). The diagnostics are 
         described in Figure \ref{fig:diag}, with the azimuthal current 
         density $j_\phi\in(-189,365)$ and the current density in the 
         $(r,z)$ plane $j_r\in(-30,109)$ and $j_z\in(-120,106)$. 
         The temperature variation is such that $\max|\tilde{T}|=2.88$, 
         and the measured $\max|u_r|=2.36$, $\max|u_\phi|=0.87$, 
         $\max|u_z|=2.35$, and $\max|B_\phi|=3.96$. 
         }
\label{fig:Q32om01}
\end{figure*}

\begin{figure}
\centerline{
\scalebox{0.46}
{\includegraphics{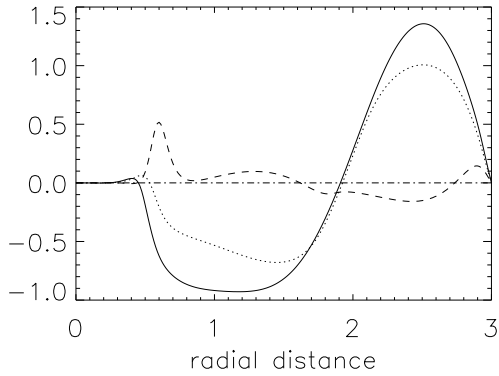}}
 }
\caption{Radial profile of $u_r$ with $\Omega=0.1$. 
         The solid line is at depth 0.25, the dotted line at 0.5, 
         and the dashed line at 0.75.
         }
\label{fig:vr}
\end{figure}

\begin{figure}
\centerline{
\scalebox{0.46}
{\includegraphics{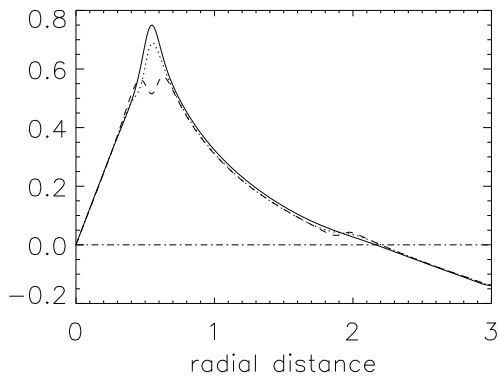}}
 }
\caption{Radial profile of azimuthal velocity $u_\phi$ with $\Omega=0.1$. 
         The lines have the same meaning as in Figure \ref{fig:vr}.
         }
\label{fig:vphi}
\end{figure}

\begin{figure}
\centerline{
\scalebox{0.46}
{\includegraphics{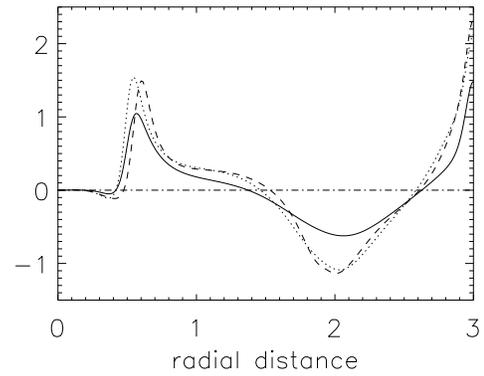}}
 }
\caption{Radial profile of the axial or vertical velocity $u_z$ with 
         $\Omega=0.1$.
         The three lines have the same meaning as in Figure \ref{fig:vr}. 
         }
\label{fig:vz}
\end{figure}

\begin{figure}
\centerline{
\scalebox{0.46}
{\includegraphics{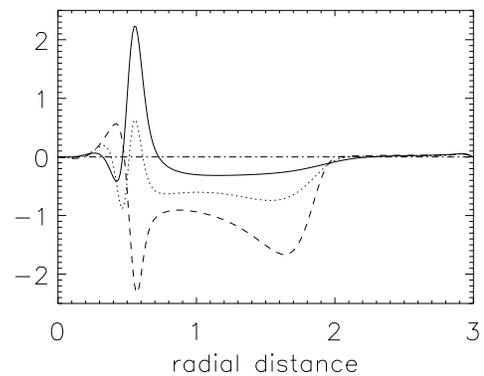}}
 }
\caption{Radial profile of azimuthal magnetic field $B_\phi$ 
         with $\Omega=0.1$.
         The three lines have the same meaning as in Figure \ref{fig:vr}. 
         }
\label{fig:bphi}
\end{figure}


\section{Numerical results}
\label{sec:numres}

Unless otherwise stated, the results shown here have been obtained with 
the following parameter values: 
$R=10^5$, $Q=32$, $\sigma=1$, $\zeta_0=0.2$, $\theta=10$, $m=1$, 
$\gamma=5/3$ and $\Gamma=3$. The results are 
presented in the format given in Figure \ref{fig:diag}. 


\subsection{No rotation}
\label{sec:norot}

When no is rotation present ($\Omega=0$ or $Ro \rightarrow \infty$), 
the azimuthal magnetic field is perturbed initially and the plasma 
is allowed to evolve through time. The solution reaches the state 
depicted in Figure \ref{fig:Q32om0}, 
with the explanation of the diagnostic given by the left hand 
box in Figure \ref{fig:diag}. The solution described here is 
typical of the plasma state when the axisymmetric cylinder is not 
rotated \citep{HurlburtRucklidge00,BothaEA06} and it provides a 
convenient base against which to compare results obtained with rotation. 

From an initial vertical magnetic field, the convection sweeps the 
magnetic field towards the central axis where it forms a flux tube. 
A large anticlockwise convection 
cell forms next to the tube, flowing towards the flux tube at the 
top of the numerical domain. This flow direction keeps the magnetic field 
confined to the central axis. The temperature is time dependent in that 
a cold plasma blob forms at the top next to the magnetic flux tube, is 
convected down the side of the tube, only to dissipate as it 
is convected along the bottom boundary. Figure \ref{fig:Q32om0} shows 
a new cold plasma blob forming at the top while the remnants of the   
previous cold blob is still visible at the bottom, moving towards the 
outer boundary. This temperature oscillation has a period of approximately 
1.275 time units, which corresponds roughly to half the circulation time 
around the convection cell. The upper layers
of the plasma are heated by the upflow next to the outer 
boundary and the resulting hot plasma blob is time independent. 
The azimuthal current density has its maximum value (in both directions) next 
to the flux tube where the magnetic field gradient is the highest.   
Azimuthal flow and magnetic structure are not generated spontaneously. 
The azimuthal components, generated by the initial perturbation, are 
small and decay exponentially as the solution evolves through time. 

Periodic oscillations, an example of which is the time dependent temperature 
next to the magnetic flux bundle, are familiar from Rayleigh-Benard 
convection \citep{CleverBusse95}.  \citet{JonesGalloway93} found 
periodic oscillations for a Boussinesq fluid in an axisymmetric cylinder. 
As must be expected, as in our case they found no spontaneous generation of 
azimuthal velocity or magnetic field components. 

Given the model's temperature and density profiles in (\ref{eq:T}) 
and (\ref{eq:rho}), the sound speed increases and the Alfv\'en speed 
decreases with depth. The inflowing layer at the top of the 
box is deeper than the outflowing layer at the bottom. This is ascribed 
to the fact that the total radial momentum in the system is zero. The 
higher density at the bottom leads to a shallower outflowing layer with 
lower radial velocities transporting the same momentum outwards as what 
the deeper top layer with higher velocities transports inwards.  
The simulation 
runs with a maximum Mach number of approximately 1. The time step is limited 
by the thermal diffusivity, with the dimensionless thermal conductivity $K$
calculated using (\ref{eq:R}). This constraint on the time step is true for 
all the numerical simulations in this paper. 


\subsection{Introduce rotation}
\label{sec:rot1}

By introducing a constant angular velocity of $\Omega=0.1$ ($Ro=77.5$) 
we obtain the solution in Figure \ref{fig:Q32om01}. 
The one convection cell in the case of no rotation (Figure \ref{fig:Q32om0}) 
has split into two cells, i.e.\ a finite $\Omega$ reduces the characteristic 
wavelength of the convection in the radial direction. This effect of 
rotation on convection is well known \citep{Chandra61}. 
The convection cell next to the flux tube always has an anticlockwise flow 
direction with an inward flow at the top, so that it forms a collar which 
forces the magnetic field together at the central axis \citep{BothaEA06}. 

By introducing a finite $\Omega$, the centrifugal term in equation 
(\ref{eq:NS}) provides a force in the radial direction. This manifests 
as a change in density contours, which go from being approximately 
horizontal without rotation (Figure \ref{fig:Q32om0}) to being slanted at 
the outer wall (Figure \ref{fig:Q32om01}). This boundary effect is localized 
and does not affect the solution in the domain interior. 
The treatment of the outer boundary is discussed in more detail in 
Section \ref{sec:out}. 

There is no time dependence in the solution. The 
radial profiles of the velocities are given in Figures \ref{fig:vr} to 
\ref{fig:vz} at three depths. All three velocity components 
are of the same order of magnitude. The radial velocity (Figure \ref{fig:vr}) 
shows the two cells circulating in opposite directions, as well as the fact 
that the speed is higher in the upper part of the numerical domain.

\begin{figure}
\centerline{
\scalebox{0.46}
{\includegraphics{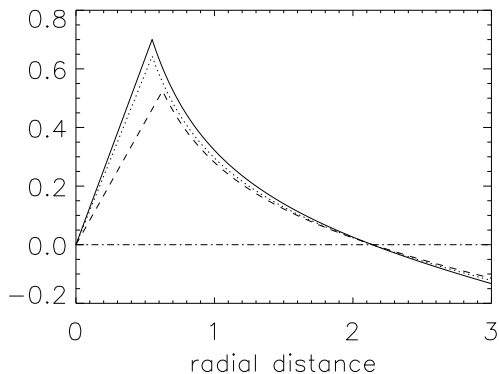}}
 }
\caption{Radial profile of a Rankine vortex inside a reference frame 
         rotating with $\Omega=0.1$. $v_\phi^\prime$ is described by 
         (\ref{eq:vmeasure}) and should be compared with Figure 
         \ref{fig:vphi}.  
         The lines have the same meaning as in Figure \ref{fig:vr}, 
         and the Rankine constants are listed in Table \ref{tab:Rankine}. 
         }
\label{fig:Rankine}
\end{figure}

The azimuthal velocity (Figure \ref{fig:vphi}) shows that the plasma 
inside the strong magnetic field of the flux tube rotates as a solid body, 
with maximum rotation on the outside edge of the tube. 
In the convection area of the solution the rotation is in the form of a 
vortex with the azimuthal velocity gradually falling away with radius. 
This rotation pattern is uniform throughout the depth of the box and 
compares well with observations that show the largest azimuthal 
velocities are located in the penumbra \citep{BrownEA03}.
One can fit the profile with a Rankine vortex, described by  
\begin{equation}
v_\phi(r)=
\left\{
\begin{array}{c@{\quad} l}
\displaystyle\frac{V_0r}{R} & \mbox{for}\quad r\leq R, \\
 & \\
\displaystyle\frac{V_0R}{r} & \mbox{for}\quad r>R,
\end{array}
\right.
\label{eq:Rankine}
\end{equation}
with $R$ the magnetic flux tube radius and $V_0=\max(u_\phi)$. 
An observer in the rotating reference frame of the cylinder will measure 
an azimuthal velocity profile of  
\begin{equation}
v_\phi^\prime(r)= v_\phi(r)-\Omega r.
\label{eq:vmeasure}
\end{equation}
The values of $V_0$ and $R$ measured for $\Omega=0.1$ are presented in 
Table \ref{tab:Rankine} and the radial profile of $v_\phi^\prime$ 
in Figure \ref{fig:Rankine}. 
Rankine vortices are used regularly to model tropical cyclones on Earth. 
Helioseismic measurements of flow around sunspots in the upper convection 
zone show a strong resemblance to the flow of hurricanes on Earth 
\citep{ZhaoKosovichev03}. It is a happy coincidence that Herschel thought 
of sunspots as large cyclonic storms \citep{ThomasWeiss92}. 
In our model the Rankine vortex makes physical sense. 
Convection is suppressed where the magnetic field is strong. 
The radial dependence of the azimuthal velocity in these regions is that  
of a rotating rigid body. Since the region experiencing rigid rotation 
corresponds to the magnetized region in all cases, we deduce that 
magnetic effects are responsible for the rigid body rotation. 
A vortex exists around the flux tube. Angular momentum mixes in 
axisymmetric convection, which results in the free vortex in the field-free 
convection cells where convection is strong. The counter flow near the 
outer wall is a consequence of the treatment of $L_z$ in our solution. 
Since our solution has zero vertical angular momentum relative to the 
rotating reference frame, a significant counter flow has to occur at the 
edge in order to balance the peak flow next to the flux tube. 

The vertical velocity (Figure \ref{fig:vz}) shows the strong downflow 
at the outside of the magnetic flux tube and at the outer edge of the 
numerical domain, as well as the upflow between the two convection cells.  
Comparing the two downflows, we observe that the downflow at the
outer edge is stronger than that next to the magnetic flux tube. 
It is essential to use a large enough aspect ratio ($\Gamma$) so 
that the outer boundary is removed from the physics around the 
magnetic flux tube. A $\Gamma=3$ appears to be a reasonable compromise 
between this and the computational limitations. 

The radial and axial magnetic field components are concentrated in 
the magnetic flux tube, with $B_z$ three times larger than $B_r$. 
Figure \ref{fig:bphi} shows the radial profile of $B_\phi$, the size of 
which is an order of magnitude smaller than $B_r$. $B_\phi$ is confined 
mainly to the inner convection cell next to the magnetic flux tube. 
The current, obtained from the magnetic field through 
equation (\ref{eq:aux}), reflects the distribution of the magnetic field. 
Its azimuthal component is concentrated on the outside of the magnetic 
flux tube where the radial gradient in the magnetic field is the largest. 
The radial and vertical components are distributed in and around the inner 
convection cell around the azimuthal magnetic field maxima (Figure 
\ref{fig:Q32om01}). At the top and bottom boundaries the radial current 
density has local maxima due to the fact that no current flows out of the box. 


\begin{table}
 \centering
 \begin{minipage}{140mm}
  \caption{The measured constants in the Rankine vortex (\ref{eq:Rankine}).}
  \label{tab:Rankine}
  \begin{tabular}{@{}cccc@{}}
  \hline
   $\Omega$ &  $z$ (measured downward) & $V_0$ & $R$  \\
  \hline
  0.1 & 0.25 & 0.75 & 0.55 \\
      & 0.5  & 0.69 & 0.55 \\
      & 0.75 & 0.57 & 0.63 \\
  \hline
  0.2 & 0.25 & 1.16 & 0.68 \\
      & 0.5  & 1.09 & 0.68 \\
      & 0.75 & 0.95 & 0.73 \\
  \hline
  0.3 & 0.25 & 1.27 & 0.85 \\
      & 0.5  & 1.24 & 0.84 \\
      & 0.75 & 1.13 & 0.88 \\
  \hline
  \end{tabular}
 \end{minipage}
\end{table}


\subsection{Increase rotation}
\label{sec:rot2}

The convective Rossby number ($Ro$) associated with the rotation around 
the central axis decreases as $\Omega$ increases. 
The $Ro$ associated with the circulation around the convection 
cells for the case with $\Omega=0.1$, i.e.\  $Ro=77.5$, 
compares well with that of supergranulation in the Sun. 
$Ro$ of larger $\Omega$ values  
correspond to even larger-scale flows. In all cases the rotation 
rate is low enough that it should not significantly change the value of the 
critical Rayleigh number for the onset of convection (see 
\citet{BrummellEA96}) and thus the cases exhibit comparable amplitudes. 

\begin{figure*}
\centerline{
\begin{minipage}{17cm}
\begin{minipage}{8cm}
\scalebox{0.4}
{\includegraphics{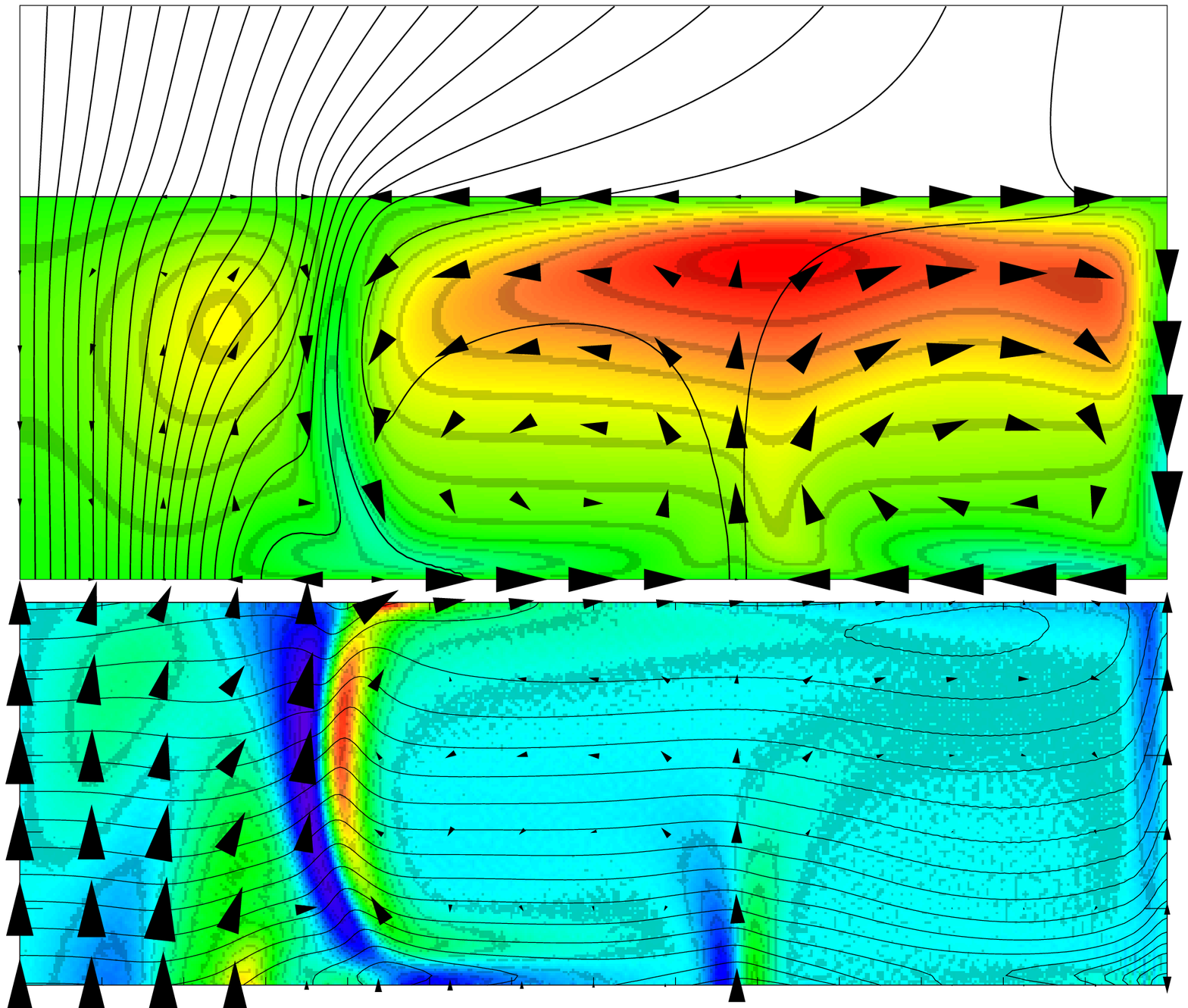}}
\end{minipage}
\begin{minipage}{8cm}
\scalebox{0.4}
{\includegraphics{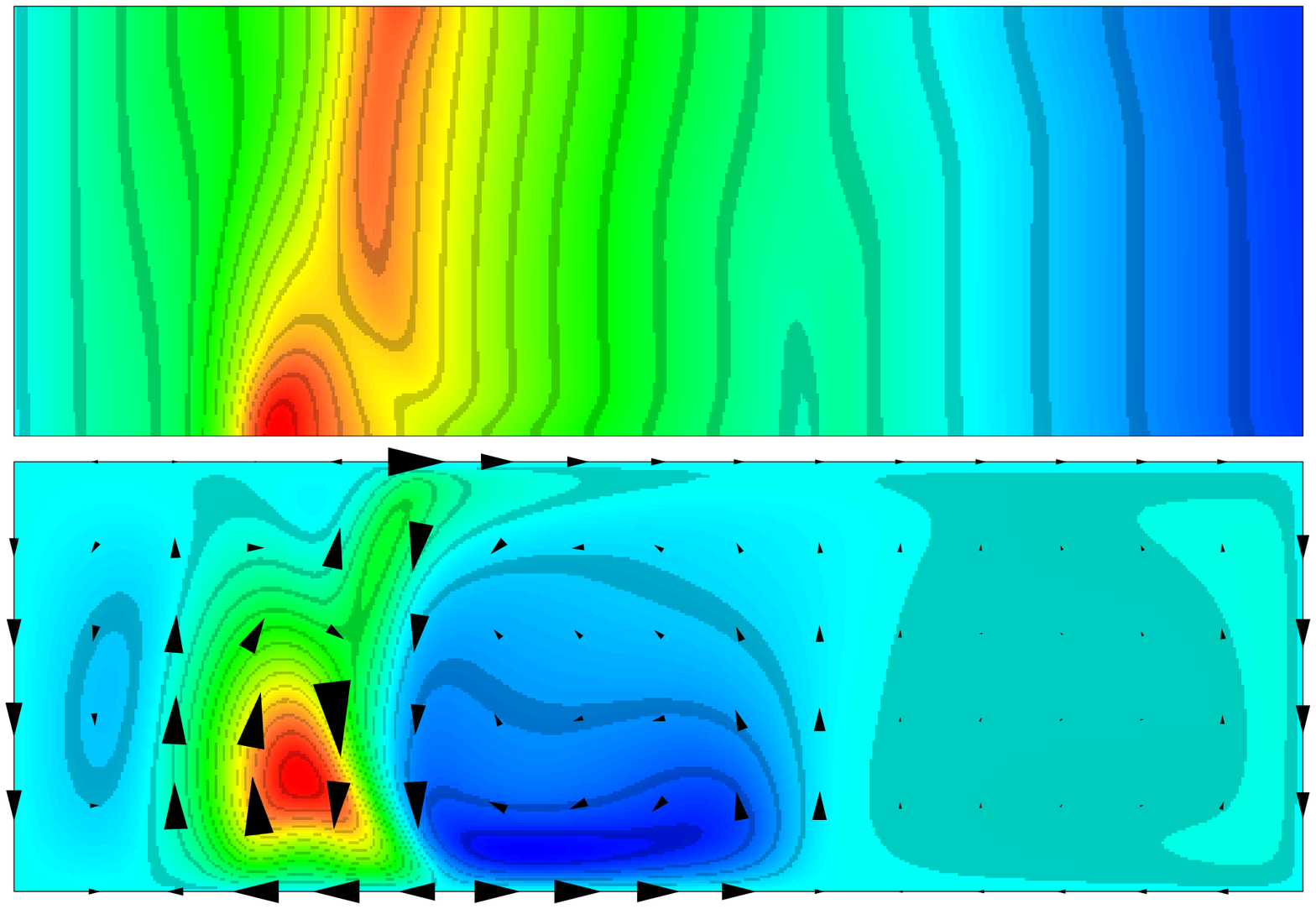}}
\end{minipage}
\end{minipage}
}
\caption{Constant angular velocity $\Omega=0.3$. 
         The convection inside the magnetic flux tube is stronger  
         and the flux tube wider than in th case with 
         $\Omega=0.1$ (Figure \ref{fig:Q32om01}). 
         The growth in width is at the expense of the inner convection 
         cell. The weak convection creates a temperature signature inside
         the flux tube. 
         The measured $\max|u_\phi|=1.5$, $\max|B_\phi|=8.0$, 
         $\max|\tilde{T}|=2.9$, $j_\phi\in(-143,220)$, 
         $j_r\in(-61,75)$, and $j_z\in(-201,111)$.
         }
\label{fig:Q32om03}
\end{figure*}

As the magnitude of $\Omega$ increases, the width of the magnetic flux 
tube increases. Figure \ref{fig:Q32om03} shows a time independent solution 
with $\Omega=0.3$. The magnetic field strength 
inside the flux tube decreases with increasing width, allowing weak 
convection to form inside the flux tube itself. 
Figure \ref{fig:Q32om03} shows that the upflow 
in the flux tube heats the plasma in the top layers of the tube, while 
very weak outflow forms along the top boundary. 
Eventually, for $\Omega\geq 0.3$, the convection inside the flux tube 
becomes strong enough to break it into concentric rings.

As rotation increases and with it the width of the magnetic flux tube, 
the magnetic field lines forming the flux tube straighten. This causes 
the azimuthal current density $j_\phi$ to decrease in both positive and 
negative azimuthal directions. The position of $j_\phi$ stays the same: 
it flows around the flux tube, created by large magnetic field gradients 
there.  

Increasing $\Omega$ also increases the size of the centrifugal force 
in equation (\ref{eq:NS}). For $\Omega= 0.1$ the density contours 
inside the flux tube are approximately horizontal (Figure 
\ref{fig:Q32om01}). For $\Omega\geq 0.2$ density contours become 
slanted, due to  
weak convection inside the flux tube as well as the centrifugal force 
acting on the plasma. As a result, there is a slight depletion of plasma 
at the top of the magnetic flux tube near the central axis, while the area 
of density variation along the inside edge of the flux tube increases. 
Figure \ref{fig:Q32om03} shows that inside the convection cells the 
increase in $\Omega$ causes a slight depression of density contours. 


\begin{figure}
\centerline{
\scalebox{0.46}
{\includegraphics{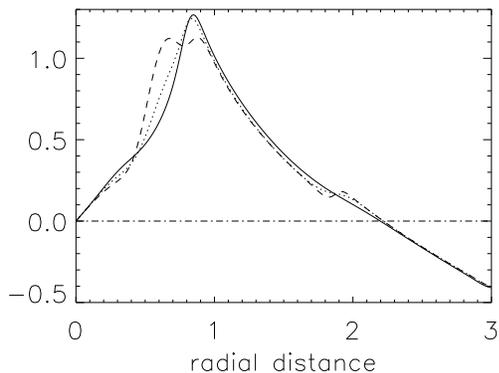}}
 }
\caption{Radial profile of azimuthal velocity $u_\phi$ with $\Omega=0.3$, 
         obtained from Figure \ref{fig:Q32om03}. 
         The lines have the same meaning as in Figure \ref{fig:vr}. 
         }
\label{fig:vom03}
\end{figure}

The plasma rotates as a Rankine vortex for all values of $\Omega$. 
Where the magnetic field is strong enough to suppress convection, 
the flow is a forced vortex in the form of rigid body rotation, 
while in the field-free convection region we observe a free vortex. 
The $1/r$ dependence in the convection region corresponds to homogeneous 
angular momentum that is caused by the effective mixing by the convection. 
As the width of the magnetic flux tube increases 
with the increase in $\Omega$, the radius of the forced vortex also 
increases (Table \ref{tab:Rankine}). The radial profiles of $v_\phi$ for 
different $\Omega$ values can be compared in Figures \ref{fig:vphi} 
and \ref{fig:vom03}. They show that as $\Omega$ increases, 
the maxima next to the magnetic flux tubes increase as well. (See also 
Table \ref{tab:Rankine}.) 
To maintain the initial $L_z=0$, the counterflow at the outer wall 
increases in sympathy. 
This is in contrast to \citet{JonesGalloway93}, who found a retrograde 
flow at the central axis for large $Q$ values in a Boussinesq fluid. 
To generate a retrograde flow near the central axis, we had to change 
the temperature boundary condition at the lower boundary in our model 
(Section \ref{sec:Tflux}). 
The rigid body 
rotation inside the magnetic flux tube is perturbed when the weak 
convection inside the tube becomes strong enough to influence the 
local magnetic field. Figure \ref{fig:Q32om03} shows 
the strength of the convection inside the flux tube, while Figure 
\ref{fig:vom03} shows the deviation from rigid body rotation. This 
deviation increases deeper in the numerical domain where convection 
is stronger.  


\begin{figure}
\centerline{
\scalebox{0.49}
{\includegraphics{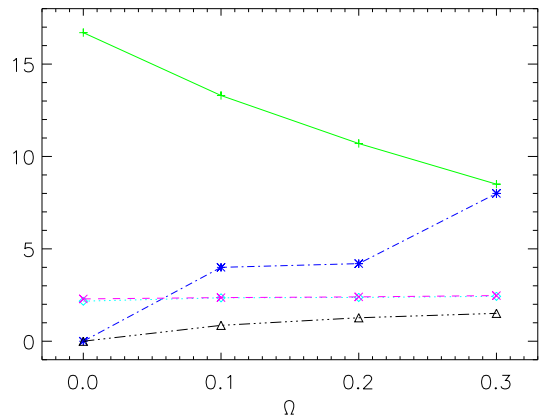}}
 }
\caption{The behaviour with increasing rotation is presented using the 
         following notation: 
         $B_r$ is green plus signs connected by a solid line; 
         $B_\phi$ is blue stars connected by a dot-dashed line; 
         $u_z$ is cyan diamonds connected by a dotted line; 
         $u_r$ is magenta crosses connected by a dashed line; 
         $u_\phi$ is black triangles connected by a triple-dot-dashed line.
         The peak values are plotted in each case. }
\label{fig:summary}
\end{figure}


\begin{figure*}
\centerline{
\begin{minipage}{17cm}
\begin{minipage}{8cm}
\scalebox{0.4}
{\includegraphics{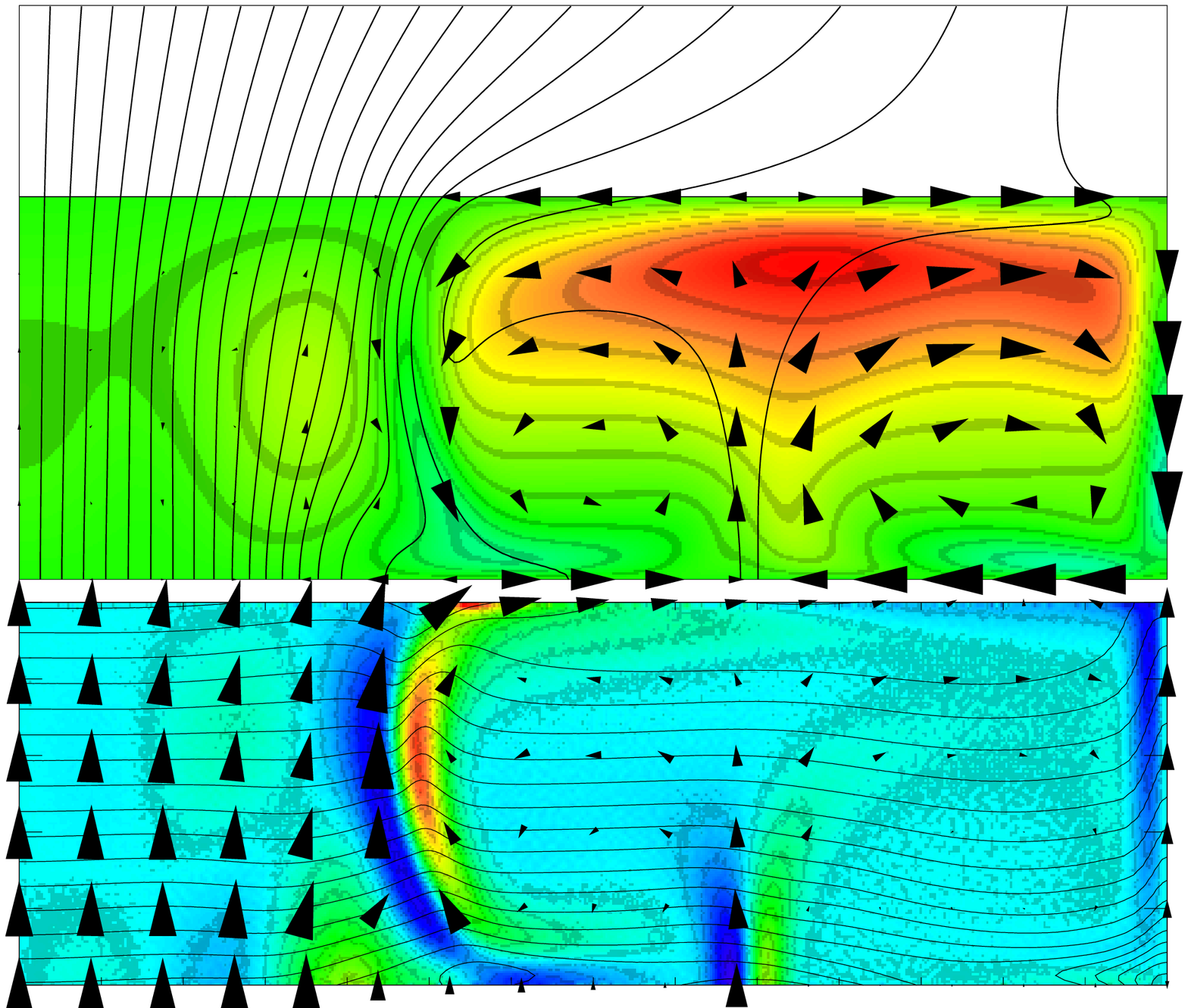}}
\end{minipage}
\begin{minipage}{8cm}
\scalebox{0.4}
{\includegraphics{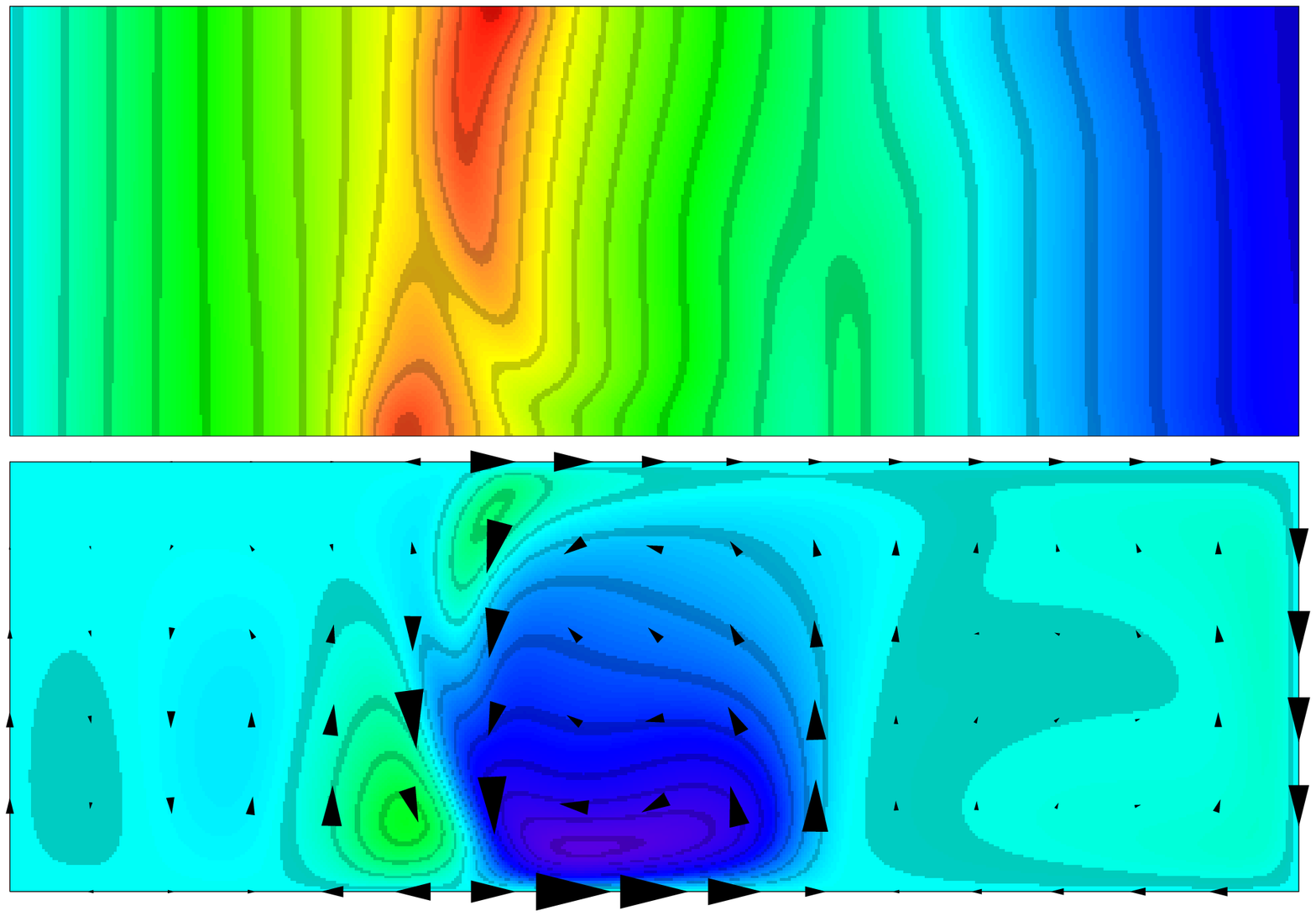}}
\end{minipage}
\end{minipage}
}
\caption{Results with $Q=128$ and $\Omega=0.3$. 
         The stronger magnetic field widens the magnetic flux tube 
         at the expense of the size of the convection area, 
         as seen when compared to Figure \ref{fig:Q32om03}. The strength  
         of convection and the size of the azimuthal quantities are  
         reduced. Max$|B_\phi|$ is located in the anticlockwise 
         convection cell next to the flux tube. 
         The measured $\max|u_\phi|=1.0$, $\max|B_\phi|=2.9$, 
         $\max|\tilde{T}|=2.7$ and $j_\phi\in(-65,125)$.
         The range of the current density in the $(r,z)$ plane is 
         $j_r\in(-13,54)$ and $j_z\in(-74,33)$.
         }
\label{fig:Q128om03}
\end{figure*}

\begin{figure}
\centerline{
\scalebox{0.46}
{\includegraphics{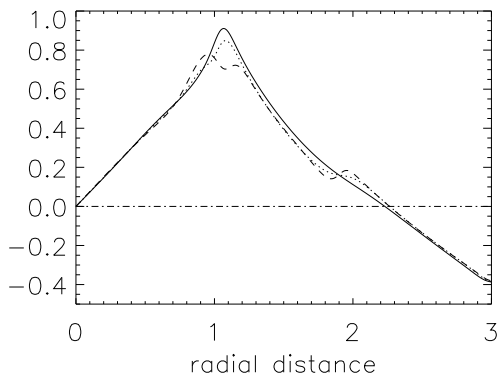}}
 }
\caption{Radial profile of $u_\phi$ corresponding to Figure 
         \ref{fig:Q128om03}, with $\Omega=0.3$ and $Q=128$.
         The lines have the same meaning as in Figure \ref{fig:vr}. 
         }
\label{fig:vQ128}
\end{figure}


The azimuthal magnetic field tends to be located in the convection 
cells closest to the central axis. In the case of $\Omega=0.1$ this 
is in the collar flow around the flux bundle (Figure \ref{fig:Q32om01}). 
For $\Omega\geq 0.2$ a small convection cell starts to form inside the
flux bundle at its base, due to weak convection inside the 
flux bundle (Figure \ref{fig:Q32om03}). 
As this cell grows in strength, the amplitude of $B_\phi$ located 
in it grows in strength relative to the $B_\phi$ in the collar flow 
outside the flux bundle. The directions of $B_\phi$ in the small cell 
inside the flux bundle and that of $B_\phi$ in the collar flow are 
anti-parallel. This corresponds to the direction of flow of the convection 
cell. Figure \ref{fig:Q32om03} shows that a clockwise convection cell has 
a $B_\phi$ pointing in the positive $\phi$ direction (i.e.\ into the page), 
while $B_\phi$ in an anticlockwise cell points in the negative $\phi$ 
direction (i.e.\ out of the page). This is caused 
by the interaction of the magnetic field with the velocity in the first 
term on the right hand side of equation (\ref{eq:Bphi}). The current 
surrounding the local maxima of $B_\phi$ has the same direction as the 
local convection, since it is calculated using equation (\ref{eq:aux}).

The behaviour when rotation is increased is summarized in Figure 
\ref{fig:summary}. As $\Omega$ increases the magnetic flux 
tube widens and the field lines straighten and become more vertical,  
which leads to a decrease in the radial component of the magnetic field. 
At the same time the azimuthal velocity and magnetic field components 
increase, being driven by $\Omega$. Compared to these changes, the radial 
and axial velocity components stay relatively stable. All velocity 
components increase in absolute value as $\Omega$ increases. 


\subsection{Increase magnetic field strength}
\label{sec:mag}

From previous numerical studies it is known that an increase in 
the magnetic field increases the width of the magnetic flux tube 
for nonrotating solutions \citep{HurlburtRucklidge00}. 
This is also true when rotation is present, which can be seen 
when Figure \ref{fig:Q32om03} with $\Omega=0.3$ and $Q=32$ is 
compared with Figure \ref{fig:Q128om03} for $\Omega=0.3$ and $Q=128$.
The solution is time independent and the growth in flux tube width 
takes place at the expense of the radial dimensions of the convection 
cells. 

The magnetic flux tube retains its configuration, with an anticlockwise 
convection cell holding the flux tube in place.  
The stronger magnetic field suppresses the weak convection inside 
the magnetic flux tube, which was present when $Q=32$ (Figure 
\ref{fig:Q32om03}).  

As the area with strong magnetic field becomes wider, the field-free 
convective region is compressed. The decrease in the field-free area is 
accompanied by lower flow velocities of convection. Here the maximum 
Mach number of the solution is 0.8, while $\max\mbox{(Mach)}= 0.9$ 
for $Q=32$. The maximum measured azimuthal velocity for $Q=128$ 
(Figure \ref{fig:Q128om03}) is also 2/3 of what it is for $Q=32$ 
(Figure \ref{fig:Q32om03}). 

The weaker flow in the convection cell around the magnetic flux tube 
means the field lines are less compressed when compared to the case
when $Q=32$ (Figure \ref{fig:Q32om03}). This leads to lower gradients 
at the flux tube's edge, which in turn implies a lower azimuthal current 
density flowing around the flux tube, since $j_\phi$ is calculated using 
equation (\ref{eq:aux}). 

The azimuthal flow of this solution fits that of a Rankine vortex. By 
suppressing the weak convection inside the flux tube that is present 
for $Q=32$, the plasma flow inside the flux tube becomes more like 
rigid body rotation. (Compare Figures \ref{fig:vom03} and \ref{fig:vQ128}.)
The maximum $u_\phi$ next to the flux bundle is lower for $Q=128$ 
due to the lower levels of convection in the solution. This is also 
true for the counter flow at the outer wall. 

The azimuthal magnetic field has its maximum in the convection 
cells closest to the central axis.   
In Figure \ref{fig:Q32om03} with $Q=32$, there exists a small 
clockwise cell at the base of the flux tube. This cell is strong 
enough to contain a significant part of $B_\phi$, with the 
anticlockwise collar flow containing an anti-parallel $B_\phi$ 
component. Here, for $Q=128$ (Figure \ref{fig:Q128om03}), the 
small clockwise cell inside the flux tube is suppressed, so that 
$\max|B_\phi|$ is mainly located in the anticlockwise collar flow. 


\begin{figure*}
\centerline{
\begin{minipage}{17cm}
\begin{minipage}{8cm}
\scalebox{0.4}
{\includegraphics{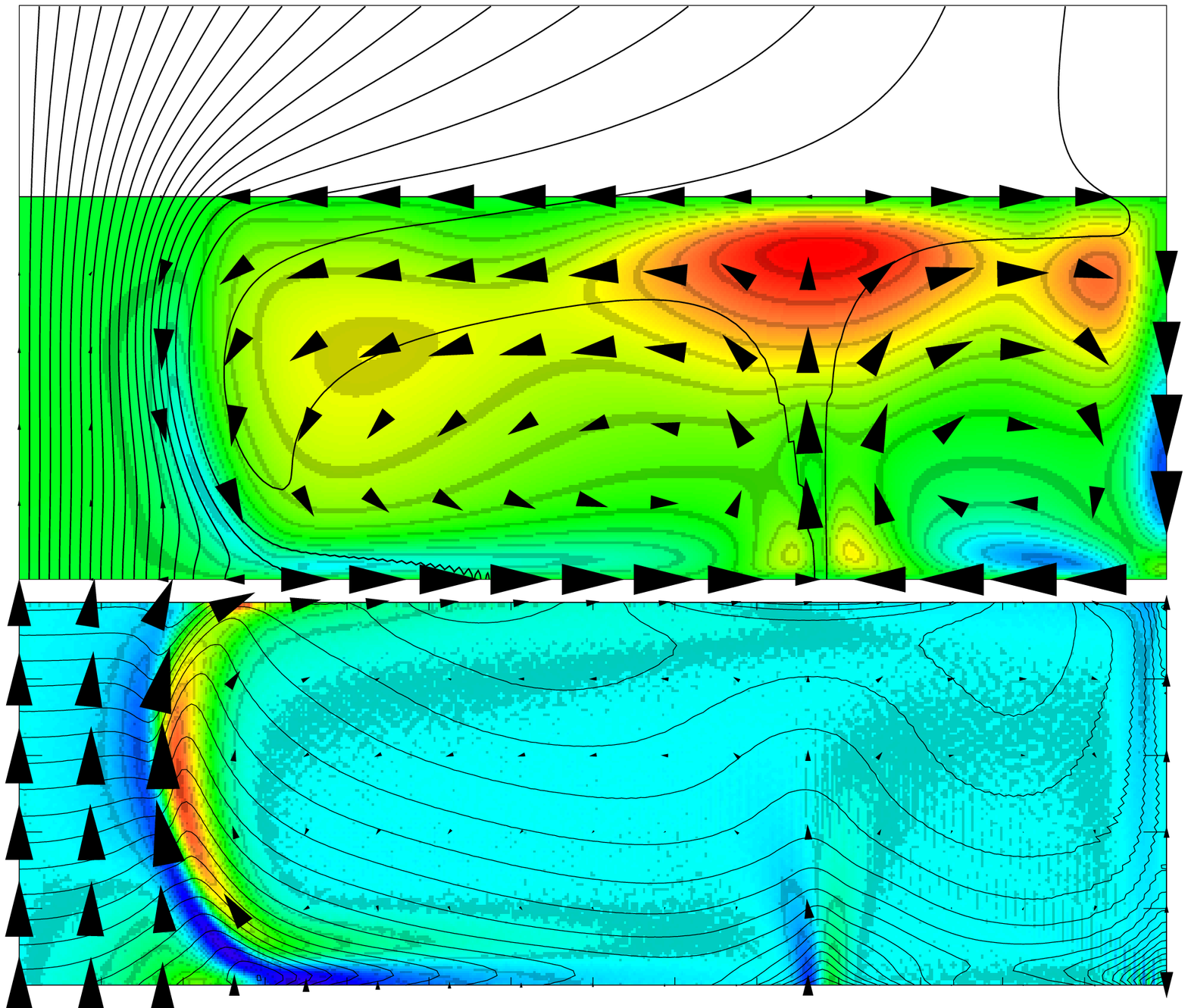}}
\end{minipage}
\begin{minipage}{8cm}
\scalebox{0.4}
{\includegraphics{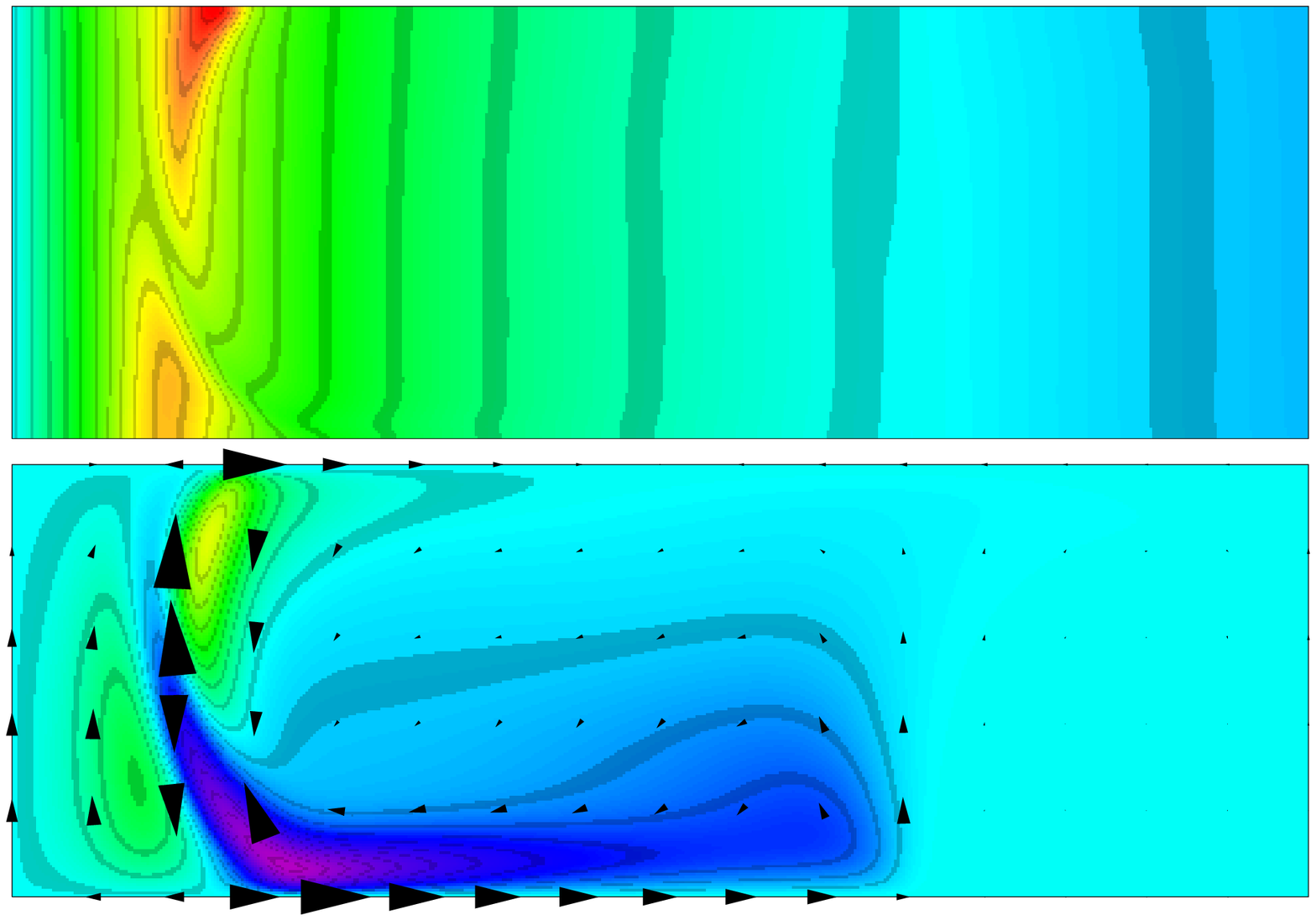}}
\end{minipage}
\end{minipage}
}
\caption{The time independent solution with $\sigma=0.3$, $Q=32$ and 
         $\Omega=0.1$.
         The lower $\sigma$ value enhances convection, which 
         leads to a larger inner convection cell, a narrower 
         central magnetic flux tube and more variation in the 
         density contours. 
         The measured $\max|u_\phi|=1.6$, $\max|B_\phi|=5.5$, 
         $\max|\tilde{T}|=3.7$, $j_\phi\in(-229,445)$, 
         $j_r\in(-49,168)$ and $j_z\in(-196,158)$.
         }
\label{fig:Q32om01sigma03}
\end{figure*}

\begin{figure}
\centerline{
\scalebox{0.46}
{\includegraphics{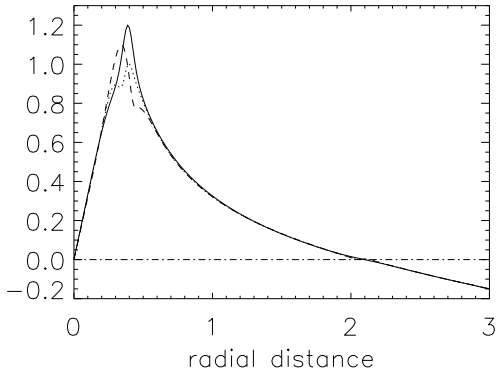}}
 }
\caption{Radial profile of $u_\phi$ corresponding to Figure 
         \ref{fig:Q32om01sigma03}, with $\Omega=0.1$, $\sigma=0.3$ 
         and $Q=32$.
         The lines have the same meaning as in Figure \ref{fig:vr}. 
         }
\label{fig:vQ32sigma03}
\end{figure}


\subsection{Lower Prandtl number}
\label{sec:Prandtl}

Decreasing the Prandtl number $\sigma$ causes the convection 
described by the steady solution to become more vigorous, as can be 
seen when Figure \ref{fig:Q32om01} (with $\sigma=1$) is compared 
to Figure \ref{fig:Q32om01sigma03} (with $\sigma=0.3$). 
For $\sigma=1$ the maximum Mach number in the solution is 0.9, 
while for $\sigma=0.3$ we have $\max\mbox{(Mach)}=1.7$.  
The lower $\sigma$ value brings the simulation closer to the 
physical conditions in the upper convection zone. The dimensionless 
thermal conductivity $K$, defined by (\ref{eq:R}), changes from 
$4.9\times10^{-2}$ for $\sigma=1$ to $1.5\times10^{-1}$ for $\sigma=0.1$. 
For $\sigma=0.3$ we have $K=8.9\times10^{-2}$ and with $\Omega=0.1$ 
we have a convective Rossby number of $Ro=42.4$.

For lower Prandtl numbers the inner convection cell increases its size 
at the expense of the width of the magnetic flux bundle, and to a lesser 
degree the width of the convection cell next to the outer boundary. Not 
only is the inner cell larger, but the velocity amplitudes  
have higher maximum values. However, the relative differences between the 
velocity components are independent of the value of $\sigma$, with the 
azimuthal component approximately a third the size of the radial and axial 
components. 
The azimuthal flow takes the form of a Rankine vortex (Figure 
\ref{fig:vQ32sigma03}). The rigid body rotation inside the flux bundle 
is faster than when $\sigma=1$ (Figure \ref{fig:vphi}), with a 
higher maximum next to the bundle. The flux bundle is also narrower, 
which means that to maintain the initial $L_z=0$ during the simulation, 
the counter flow at the outer boundary needs to be only slightly larger 
than when $\sigma=1$. 

The stronger convection pushes the magnetic flux into a thinner flux bundle 
at the central axis, so that the strength of $B_z$ increases for 
lower values of $\sigma$. The size of $B_\phi$ relative to $B_r$ stays 
approximately the same for all values of $\sigma$. As in the case for 
$\sigma=1$ (Figure \ref{fig:Q32om01}), the azimuthal magnetic field is 
mostly located in the inner convection cell, with its maxima next to the 
magnetic flux bundle. 
The stronger convection also causes the curvature and gradients of the 
magnetic field lines in the $(r,z)$ plane to increase, which increases 
the size of the azimuthal current density obtained from equation 
(\ref{eq:aux}). The position of $\max|j_\phi|$ stays the same.

The effects of the enhanced convection are visible in the density contours. 
For $\sigma=1$ the density contours in the convection cells are approximately 
horizontal (Figure \ref{fig:Q32om01}) while for $\sigma=0.3$ significant 
variation in the radial direction occurs (Figure 
\ref{fig:Q32om01sigma03}). At the central axis, where the convection is 
suppressed by the strong magnetic field, the contour lines stay 
approximately as they were with $\sigma=1$ and its lower convection strengths. 

There are changes in the temperature profile of the solution.
The stronger upflow between the two convection cells leads to a larger 
variation from the original heat profile in the upper layers of the 
solution, as can be seen when $\max|\tilde{T}|$ in Figures 
\ref{fig:Q32om01} and \ref{fig:Q32om01sigma03} are compared. 
Also, for lower $\sigma$ values the thermal diffusion rate becomes more 
significant, which reduces the radial extent of the heated plasma above the 
upflow. 


\begin{figure*}
\centerline{
\begin{minipage}{17cm}
\begin{minipage}{8cm}
\scalebox{0.4}
{\includegraphics{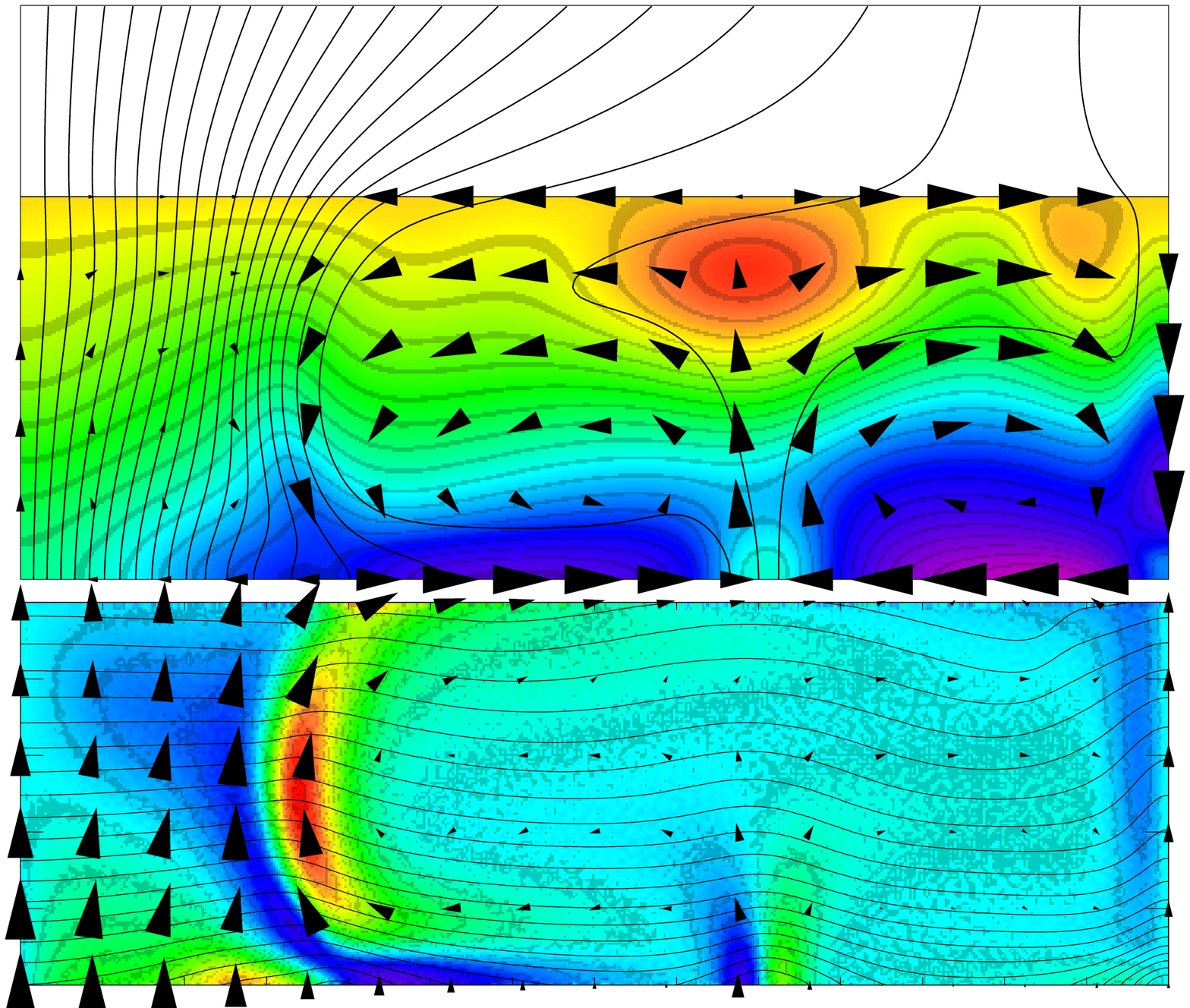}}
\end{minipage}
\begin{minipage}{8cm}
\scalebox{0.4}
{\includegraphics{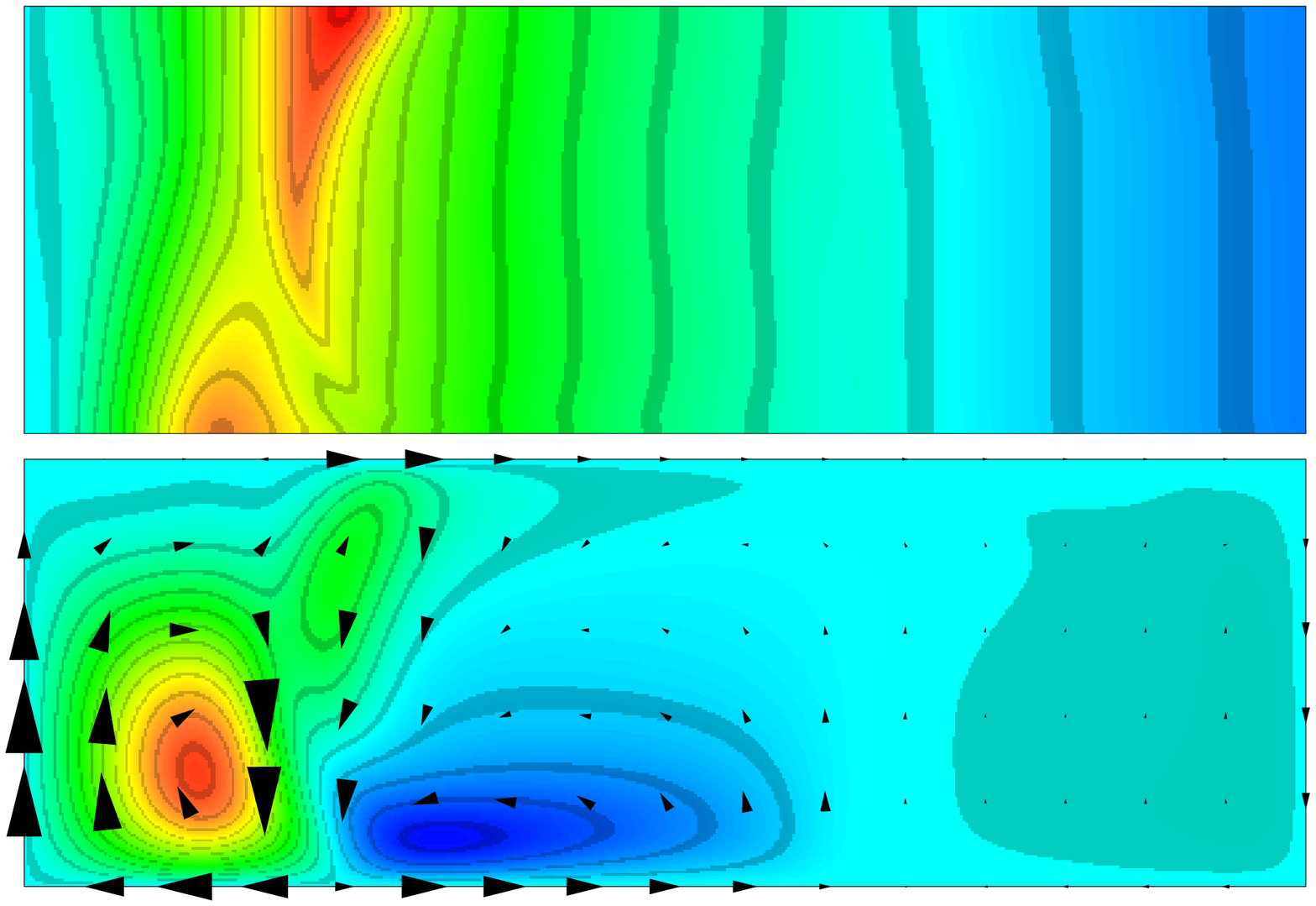}}
\end{minipage}
\end{minipage}
}
\caption{Results with a constant $\partial T/\partial z$ used as bottom 
         boundary 
         condition. The parameters are $\Omega=0.1$, $Q=32$ and $\sigma=0.1$. 
         Weak convection forms inside the wide magnetic flux tube.
         The overall convection is lower and the inner convection 
         cell smaller compared to solutions using a constant 
         $T$ lower boundary condition. 
         The measured $\max|u_\phi|=0.8$, $\max|B_\phi|=3.2$, 
         $\max|\tilde{T}|=1.4$ and $j_\phi\in(-73,105)$. 
         The current density in the $(r,z)$ plane is $j_r\in(-27,23)$ and  
         $j_z\in(-53,50)$.
         }
\label{fig:Q32om01sigma01heat}
\end{figure*}


\begin{table*}
 \centering
 \begin{minipage}{140mm}
  \caption{Changing the lower temperature boundary condition 
           with $Q=32$, $m=1$, $\Gamma=3$, $\theta=10$.}
  \label{tab:T}
  \begin{tabular}{@{}lccccc@{}}
  \hline
   $\Omega$ &  $\sigma$ & \multicolumn{2}{c}{Lower boundary: $T$ constant} &
   \multicolumn{2}{c}{Lower boundary: $\partial T/\partial z$ constant}\\
            &            &  max(Mach) & $T$ at $z=1$ & 
                            max(Mach) & $T$ range at $z=1$ \\
  \hline
  0   & 1.0 & 1.2 & 11 & 0.6 & (9.34,10.19) \\
  0.1 & 1.0 & 0.9 & 11 & 0.6 & (9.33,10.42) \\
  0.1 & 0.3 & 1.7 & 11 & 1.1 & (9.05,10.43) \\
  0.1 & 0.1 & --  & -- & 1.3 & (8.59,10.20) \\
  \hline
  \end{tabular}
 \end{minipage}
\end{table*}


\subsection{Temperature prescription at bottom boundary}
\label{sec:Tflux}

To determine the extent to which the bottom boundary is 
influencing the result, we changed the temperature prescription on 
this boundary from a constant value $T$ to a constant 
$\partial T/\partial z$. From equation (\ref{eq:T}) we set 
$\partial T/\partial z=\theta$, so that the heat flux 
$K\theta$ stays equivalent to the heat flux for a constant $T$ 
boundary condition.  
A linear stability analysis by \citet{HurlburtEA84} determined that this 
change in boundary condition results in halving the critical Rayleigh 
number for the onset of convection, and hence one would expect somewhat 
more vigorous convection for the same Rayleigh number, all other aspects 
of the solution being equal. 
However, the numerical results discussed in this section show that 
for this highly nonlinear system, the change in the lower boundary 
condition leads to lower convection levels. 

For low Prandtl numbers, such as $\sigma=0.1$ in Figure 
\ref{fig:Q32om01sigma01heat}, the basic configuration of two convection 
cells and a central magnetic flux tube remains as before. The radial profile 
of the azimuthal flow is that of a Rankine vortex. The strength of convection 
outside the magnetic flux tube is lower than for the bottom boundary 
condition of constant temperature, while the magnetic flux tube is wider 
with very weak convection inside it. 

\begin{figure*}
\centerline{
\begin{minipage}{17cm}
\begin{minipage}{8cm}
\scalebox{0.4}
{\includegraphics{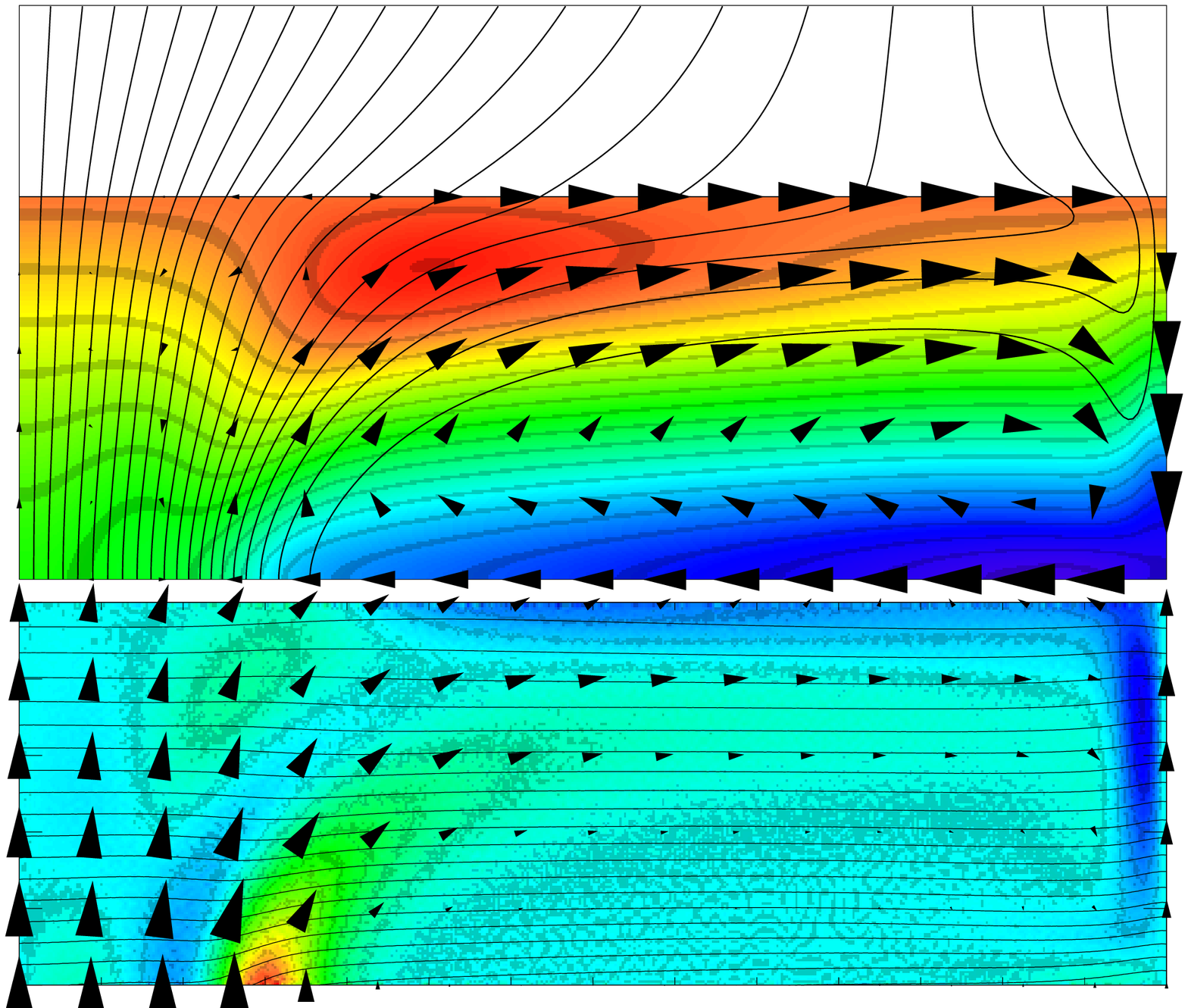}}
\end{minipage}
\begin{minipage}{8cm}
\scalebox{0.4}
{\includegraphics{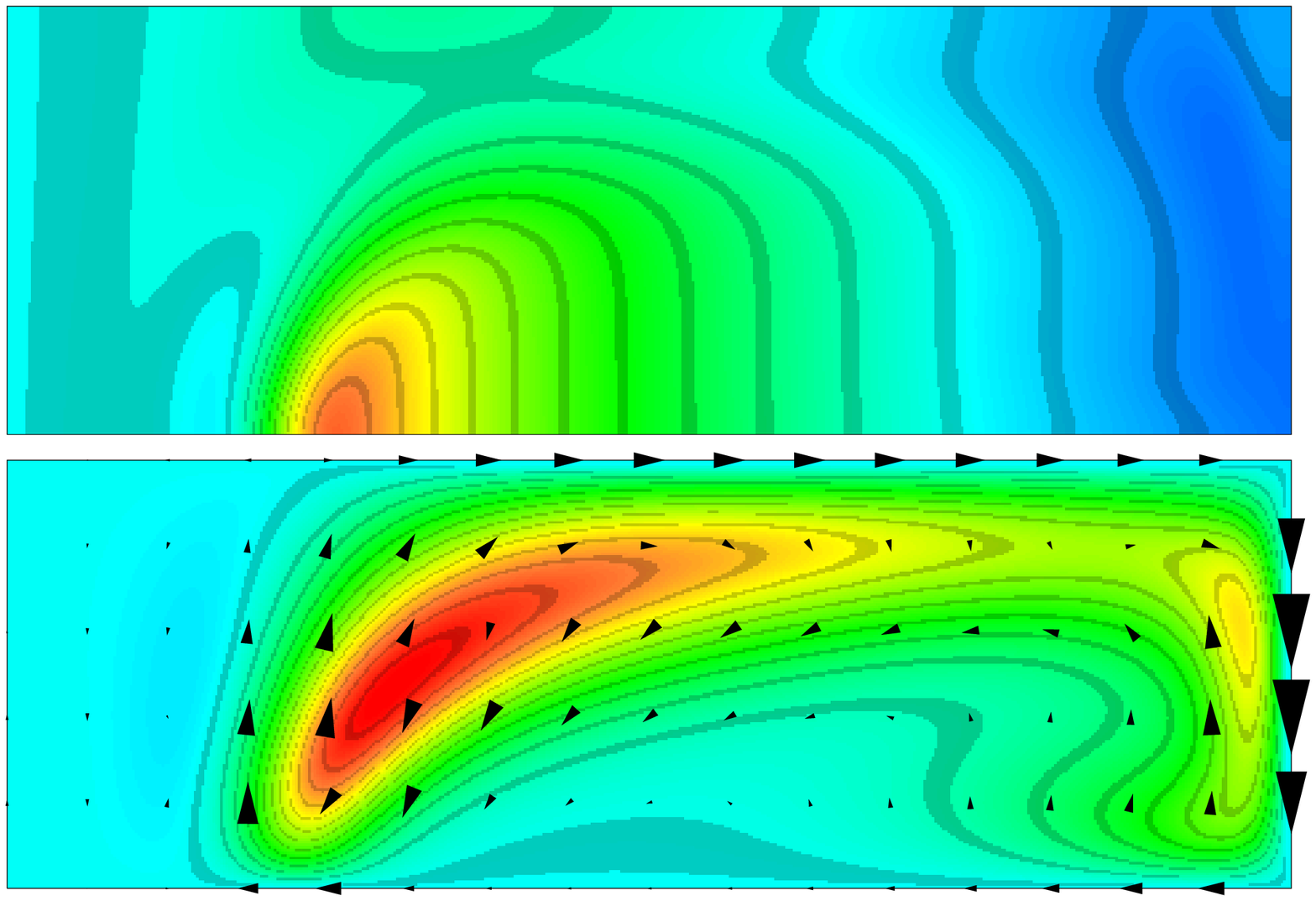}}
\end{minipage}
\end{minipage}
}
\caption{Results with constant $\partial T/\partial z$ bottom boundary 
         condition and the same parameter values as Figure 
         \ref{fig:Q32om01sigma01heat}, but with $\sigma=1$. 
         One convection cell forms with outflow along the top boundary.
         This allows the magnetic flux to spread radially, with weak 
         convection forming near the central axis inside the flux tube.
         The measured $\max|u_r|=0.8$, $\max|u_\phi|=0.5$, $\max|u_z|=0.8$, 
         $\max|B_\phi|=3.7$, $\max|\tilde{T}|=1.1$ and $j_\phi\in(-59,149)$.
         The current density in the $(r,z)$ plane is $j_r\in(-20,22)$ and  
         $j_z\in(-134,50)$.
         }
\label{fig:Q32om01heat}
\end{figure*}

\begin{figure}
\centerline{
\scalebox{0.46}
{\includegraphics{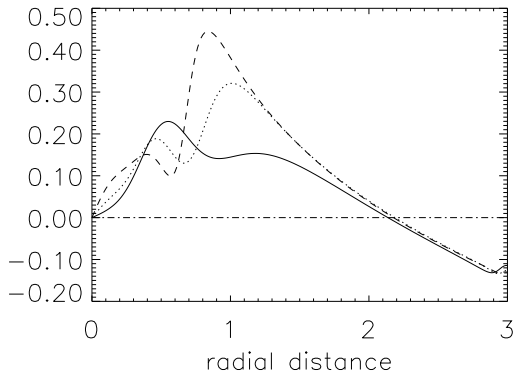}}
 }
\caption{Radial profile of $u_\phi$ with $\Omega=0.1$, $\sigma=0.3$, 
         $Q=32$ and a constant $\partial T/\partial z$ as bottom boundary.
         The solution has a configuration similar to Figure 
         \ref{fig:Q32om01heat}, 
         but the weak convection inside the flux tube has higher 
         amplitudes. These are responsible for the large perturbations 
         in the interval $0\leq r\leq 1$. 
         The lines have the same meaning as in Figure \ref{fig:vr}. 
         }
\label{fig:vQ32sigma03heat}
\end{figure}

\begin{figure}
\centerline{
\scalebox{0.46}
{\includegraphics{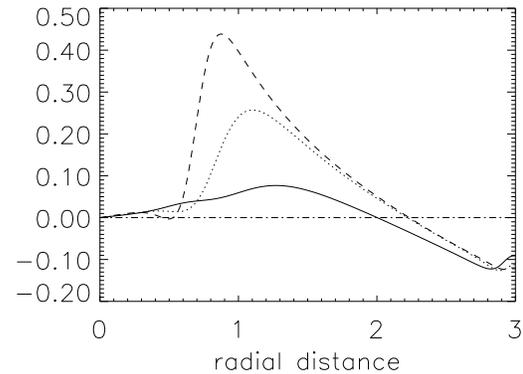}}
 }
\caption{Radial profile of $u_\phi$ corresponding to Figure 
         \ref{fig:Q32om01heat}, with $\Omega=0.1$, $\sigma=1$, 
         and $Q=32$. Figure \ref{fig:vr} defines the line notation. 
         }
\label{fig:vQ32heat}
\end{figure}

\begin{figure}
\centerline{
\scalebox{0.46}
{\includegraphics{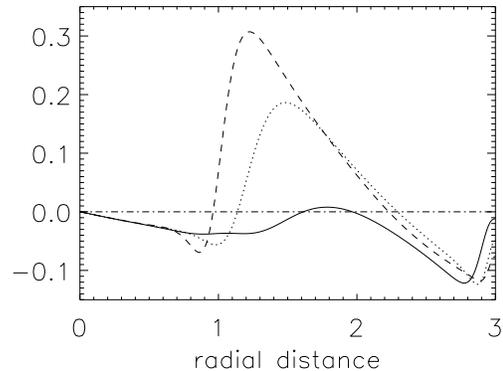}}
 }
\caption{Radial profile of $u_\phi$ with $\Omega=0.1$, $\sigma=0.3$, 
         $Q=256$ and constant $\partial T/\partial z$. 
         The solution has a configuration similar to Figure 
         \ref{fig:Q32om01heat}. 
         The lines have the same meaning as in Figure \ref{fig:vr}. 
         }
\label{fig:vQ256sigma03heat}
\end{figure}

For Prandtl numbers of $\sigma\geq 0.3$ 
the solution changes into one convection cell that is outflowing at the 
top. This new flow direction does not provide an efficient collar 
to contain the magnetic flux \citep{BothaEA06}, so that the magnetic field 
spreads out in the radial direction rather than being contained at the central 
axis. Under these circumstances it is possible to have horizontal magnetic 
field lines above the main convection cell. Figure \ref{fig:Q32om01heat} 
gives the solution for $\sigma=1$, while Figures \ref{fig:vQ32sigma03heat} 
to \ref{fig:vQ256sigma03heat} give the radial profiles of $u_\phi$ for various 
parameter values. 


Figure \ref{fig:Q32om01heat} shows the solution with $\sigma=1$. 
A solution in the same parameter space but with constant temperature at
the bottom boundary is given by Figure 
\ref{fig:Q32om01}. When comparing the two solutions, it is clear 
that the level of convection is lower in the case of constant 
$\partial T/\partial z$ boundary condition. This is shown explicitly 
in Table \ref{tab:T}: the maximum Mach number is lower when a constant 
$\partial T/\partial z$ is used. 
It also shows in the fact that the maximum azimuthal velocity is weaker 
in Figure \ref{fig:Q32om01heat} than in Figure \ref{fig:Q32om01}. 
This is true for all choices of parameter values. 
The maximum Mach number in the solution increases as $\sigma$ decreases, 
in line with the discussion in Section \ref{sec:Prandtl} and the fact 
that more heat flows through the system.  
In Table \ref{tab:T} the solution for $\sigma=0.1$ and constant 
$\partial T/\partial z$ does not have a counterpart when a constant 
temperature boundary condition is used, because the convection becomes  
too vigorous and large shocks form that terminate the numerical simulation. 
This agrees with the conclusion that a constant $\partial T/\partial z$ at 
the lower boundary leads to lower convection levels. 


The weaker convection allows the magnetic flux tube to be wider, so that 
the value of $\max|B_z|$ is lower. The different flow pattern 
when $\sigma\geq 0.3$ also contribute to the lower vertical 
magnetic component, in that the magnetic field now has a significant 
horizontal component (Figure \ref{fig:Q32om01heat}). 
Another consequence of the weaker convection is 
lower magnetic field gradients in the $(r,z)$ plane. This leads to lower 
levels of azimuthal current density, which can be seen when Figures 
\ref{fig:Q32om01} and \ref{fig:Q32om01heat} are compared. 
In the case for $\sigma=0.1$ (Figure \ref{fig:Q32om01sigma01heat})
the peak current density is positioned around the magnetic flux tube close to 
the midplane, where it is for all results with a collar flow 
around the flux tube, while for $\sigma\geq 0.3$ (Figure \ref{fig:Q32om01heat})
the maximum current density is located at the base of the flux tube, 
where magnetic field gradients are largest. 

As in the case for a constant $T$ lower boundary, the plasma rotates as a 
rigid body (or forced vortex) where the magnetic field is strongest, while 
a free vortex forms 
in the convection area where the magnetic field is weaker. For 
$\sigma\geq 0.3$ the central magnetic flux tube is still present, with 
an additional component that stretches horizontally above the dominant 
convection cell. This means that at the bottom of the numerical box, 
where one has a well defined flux tube (Figure \ref{fig:Q32om01heat}), 
the radial profile of the azimuthal flow  looks most like a Rankine vortex. 
(See Figures \ref{fig:vQ32sigma03heat} for $\sigma=0.3$ and 
\ref{fig:vQ32heat} for $\sigma=1$.) Moving higher up in the numerical domain, 
the width of the magnetic flux tube increases and with it the radius of the 
forced vortex, with the free vortex in the convection area occupying less 
space. At the top of the box the magnetic field influences the azimuthal 
flow so much that most of the free vortex flow is distorted. 
The strong convection associated with low $\sigma$ values 
(Section \ref{sec:Prandtl}) enables weak convection to form inside 
the magnetic flux tube. This weak flow serves as 
a perturbation to the rigid body rotation inside the magnetic flux tube 
around the axis (Figure \ref{fig:vQ32sigma03heat} for $\sigma=0.3$). 

By increasing $\sigma$ one weakens the convection in the solution, as 
discussed in Section \ref{sec:Prandtl}, and reduces the heat flux through 
the system. This allows the magnetic field 
to become more uniform along the central axis with less perturbations in 
the rigid body rotation that occurs there. Figure \ref{fig:Q32om01heat} 
shows an example of this for $\sigma=1$. 
An interesting phenomenon occurs when $\sigma=1$. In this case the plasma 
inside the magnetic flux tube slows down to almost zero (Figure 
\ref{fig:vQ32heat}). The plasma in the field-free convection area still 
forms a free vortex, which means the azimuthal flow grows from 
a small value to its maximum next to the flux tube over a small distance, 
before it tails off into the usual free vortex profile. The maximum flow 
in the vortex occurs close to the base of the flux tube, where the 
convection area has the lowest level of magnetic field. 
The fact that the slowly flowing plasma inside the flux tube is still in 
the same direction as the free vortex is a coincidence. The plasma flow 
inside the magnetic field area is highly sensitive to the values of $Q$ 
and $\sigma$. Some values give a retrograde rotation at the 
central axis, in the same direction as the counter flow near the outer 
wall. Other values give a prograde rotation in the direction of the 
free vortex, but with huge perturbations in the rigid body rotation, while 
others will give a flux tube that rotates partly prograde and partly 
retrograde. As one example of what can occur, we plot the radial 
profile of $u_\phi$ for the values $Q=256$ and $\sigma=0.3$ in 
Figure \ref{fig:vQ256sigma03heat}. Here the plasma in the whole of the 
flux tube rotates retrograde, as well as most of the plasma in the 
magnetic flux layer on top of the one large convection cell. (The 
solution has the same configuration as Figure \ref{fig:Q32om01heat}, 
only with the width of the flux tube wider due to the higher $Q$ 
value.) Notice in Figure \ref{fig:vQ256sigma03heat} that the free vortex 
still rotates prograde with a counter flow next to the outer wall, similar 
to Figure \ref{fig:vQ32heat} when $Q=32$ and $\sigma=1$. 

For a constant $T$ at the lower boundary, the azimuthal magnetic field 
formed in the inner convection cell closest to the central axis. Here, 
for $\sigma=0.1$ (Figure \ref{fig:Q32om01sigma01heat}), weak convection 
inside the magnetic flux tube allows a small cell to form at the 
base of the flux tube. This cell is closest to the central axis and 
carries a large part of $B_\phi$ in it. The much stronger convection 
cell forming the collar flow around the magnetic flux tube contains the 
rest of $B_\phi$. These two cells have opposite meridional circulations, 
so that the $B_\phi$ components in them are anti-parallel to each other. 
For the cases where $\sigma\geq 0.3$ (Figure \ref{fig:Q32om01heat}), 
only one cell dominate the convection area. 
The azimuthal magnetic field is located inside this cell, but with its 
maximum value next to the magnetic flux tube. It is interesting to note 
that max($B_\phi$) is situated towards the top of the numerical domain, 
while it is towards the bottom of the domain for constant $T$ lower 
boundaries. This corresponds to the direction of flow in the convection 
cell containing $B_\phi$ in each case.  

The influence of the outer wall is discernible with constant 
$\partial T/\partial z$ at the lower boundary, when only one clockwise 
convection cell forms in the solution (i.e.\ for $\sigma\geq 0.3$). 
At the outer wall a local $\max(B_\phi)$ forms, generating its own 
current around it in the $(r,z)$ plane (Figure \ref{fig:Q32om01heat}). 
The heat flux through the system with $\sigma<1$ is higher than for 
$\sigma=1$ and the convection stronger, so that the magnetic field is less 
able to concentrate next to the outer wall. One can also see the 
influence of the outside wall in the azimuthal velocity profile;  
the amplitude of the counter flow next to the outer wall  
diminishes sharply at the wall. This effect becomes larger as the size 
of the local $\max(B_\phi)$ at the outer wall increases. (Compare Figures 
\ref{fig:vQ32sigma03heat} and \ref{fig:vQ32heat} where $\sigma$ 
increases, as well as \ref{fig:vQ32sigma03heat} and 
\ref{fig:vQ256sigma03heat} where $Q$ increases.)  
In all our results this boundary effect is highly localized 
and does not influence the solution deeper in the numerical domain. 
In Section \ref{sec:out} the treatment of the outer boundary is discussed. 

\begin{figure}
\centerline{
\scalebox{0.46}
{\includegraphics{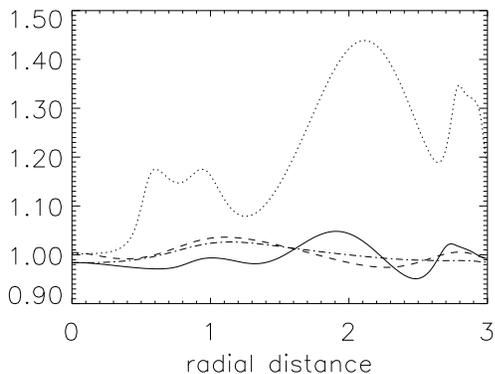}}
 }
\caption{Top boundary temperature with constant $\partial T/\partial z$ at 
         bottom boundary. The solid line is $\sigma=0.1$ 
         (Figure \ref{fig:Q32om01sigma01heat}), the broken line   
         $\sigma=0.3$ 
         and the dot-dashed line $\sigma=1$ (Figure 
         \ref{fig:Q32om01heat}). The dotted line is  
         $\sigma=0.3$ with a constant bottom temperature (Figure 
         \ref{fig:Q32om01sigma03}), added as reference. 
         }
\label{fig:heattop}
\end{figure}

\begin{figure}
\centerline{
\scalebox{0.46}
{\includegraphics{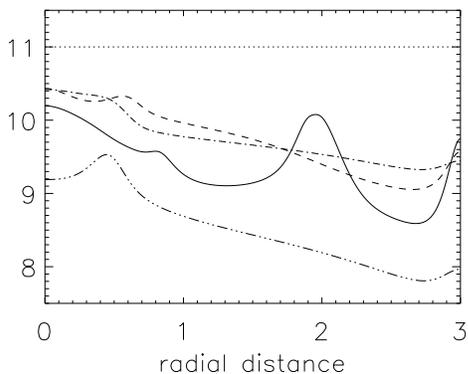}}
 }
\caption{Temperature at bottom boundary with $\partial T/\partial z$ 
         constant. The lines have the same meaning as in Figure 
         \ref{fig:heattop}. The extra (triple-dot-dashed) line 
         corresponds to $\sigma=1$ and $\theta=20$ (Figure 
         \ref{fig:Q32om01theta20heat}), as discussed in Section 
         \ref{sec:depth}. To fit into the graph, 10 has been 
         subtracted from this temperature. 
         }
\label{fig:heatbottom}
\end{figure}

Figures \ref{fig:heattop} and \ref{fig:heatbottom} show that the 
temperatures at the top and bottom domain boundaries are lower when 
constant $\partial T/\partial z$ is used, compared to a constant $T$ at 
the lower boundary. Inside the magnetic flux tube the temperature 
near the top of the domain (Figure \ref{fig:heattop}) is largely independent 
of the bottom boundary condition. Where the convection dominates outside 
the magnetic flux tube, the temperature variation is lower than 
for a constant $T$ at the bottom. Figure \ref{fig:heatbottom} shows 
the temperature variation at the lower boundary.   
Similar to the top of the domain, the 
temperature inside the magnetic flux tube is least affected by the 
boundary condition, although the temperature is lower along the whole radial 
length for $\partial T/\partial z$  constant. 
Figure \ref{fig:heatbottom} and Table \ref{tab:T} show that the variation of 
the temperature along the bottom boundary is substantial.
This observation may explain the difference between the linear 
stability results by \citet{HurlburtEA84} and the nonlinear behaviour, 
as mentioned in the beginning of this section. 
Although the heat flux into the domain stays the same, a lower temperature 
along the bottom boundary will drive the convection less vigorously. 
To investigate how sensitive the velocity amplitudes are to the 
lower bottom temperature, we increased the bottom boundary temperature and 
the heat flux through it by increasing $\theta$. This, however, has 
repercussions throughout the system, as discussed in Section \ref{sec:depth}.


\begin{table*}
 \centering
 \begin{minipage}{140mm}
  \caption{Changing the stratification of the domain ($\theta$) with 
           $\sigma=1$, $m=1$, $\Gamma=3$.}
  \label{tab:theta}
  \begin{tabular}{@{}crccccc@{}}
  \hline
   $\Omega$ &  $Q$ & Lower boundary & \multicolumn{2}{c}{$\theta=10$} &
                                      \multicolumn{2}{c}{$\theta=20$}\\
            &      &                &  max(Mach) & $T$ at $z=1$ & 
                                       max(Mach) & $T$ at $z=1$ \\
  \hline
  0.1 & 32  & $\partial T/\partial z$ constant & 0.6 & (9.33,10.42)  
                                               & 0.8 & (17.81,19.53) \\
  0.3 & 128 & $T$ constant  & 0.8 & 11 & 1.0 & 21 \\
  \hline
  \end{tabular}
 \end{minipage}
\end{table*}


\subsection{Increase stratification in numerical domain}
\label{sec:depth}

Increasing the value of $\theta$ from 10 to 20 is felt throughout the 
system. From equations (\ref{eq:T}) and (\ref{eq:rho}) we see that the 
stratification doubles. Through equation (\ref{eq:R}) the value 
of the dimensionless thermal conductivity $K$ changes from $4.9\times10^{-2}$ 
for $\theta=10$ to $1.3\times10^{-1}$ for $\theta=20$. This means 
the heat flux through the system ($K\theta$) is raised five fold, 
as mentioned in Section \ref{sec:Tflux}. It also implies that the 
convective Rossby number (\ref{eq:Ro}) increases one order of magnitude 
while $\Omega$ stays unchanged. 
Table \ref{tab:theta} shows that the maximum Mach number in the solution 
increases for both temperature prescriptions. 
The linear stability properties are changed both by increasing the 
critical Rayleigh number for the onset of convection, due to the change in 
stratification \citep{HurlburtEA84}, and by increasing the mean value of 
$\zeta$ in the domain \citep{WeissEA90}. 

\begin{figure*}
\centerline{
\begin{minipage}{17cm}
\begin{minipage}{8cm}
\scalebox{0.4}
{\includegraphics{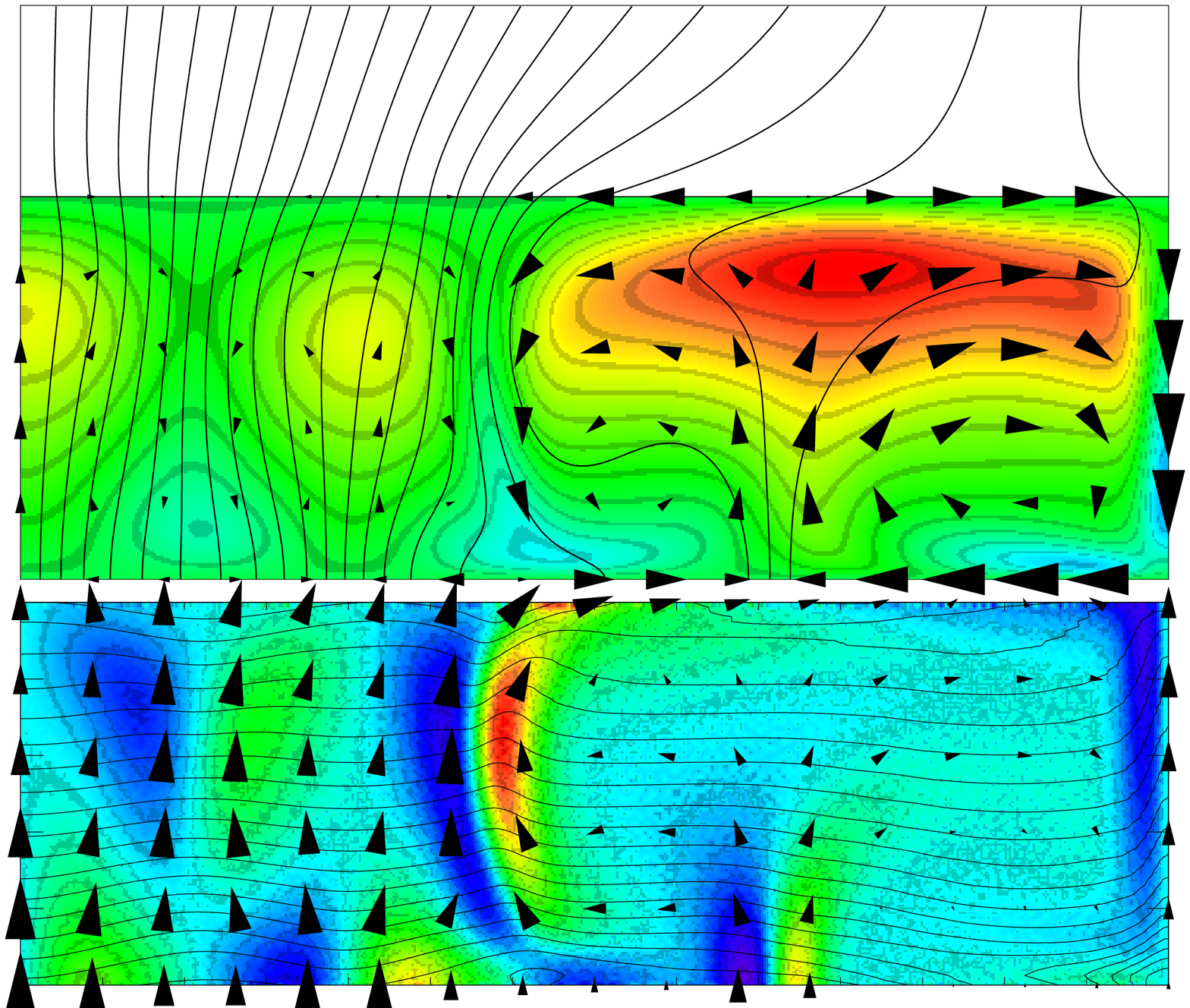}}
\end{minipage}
\begin{minipage}{8cm}
\scalebox{0.4}
{\includegraphics{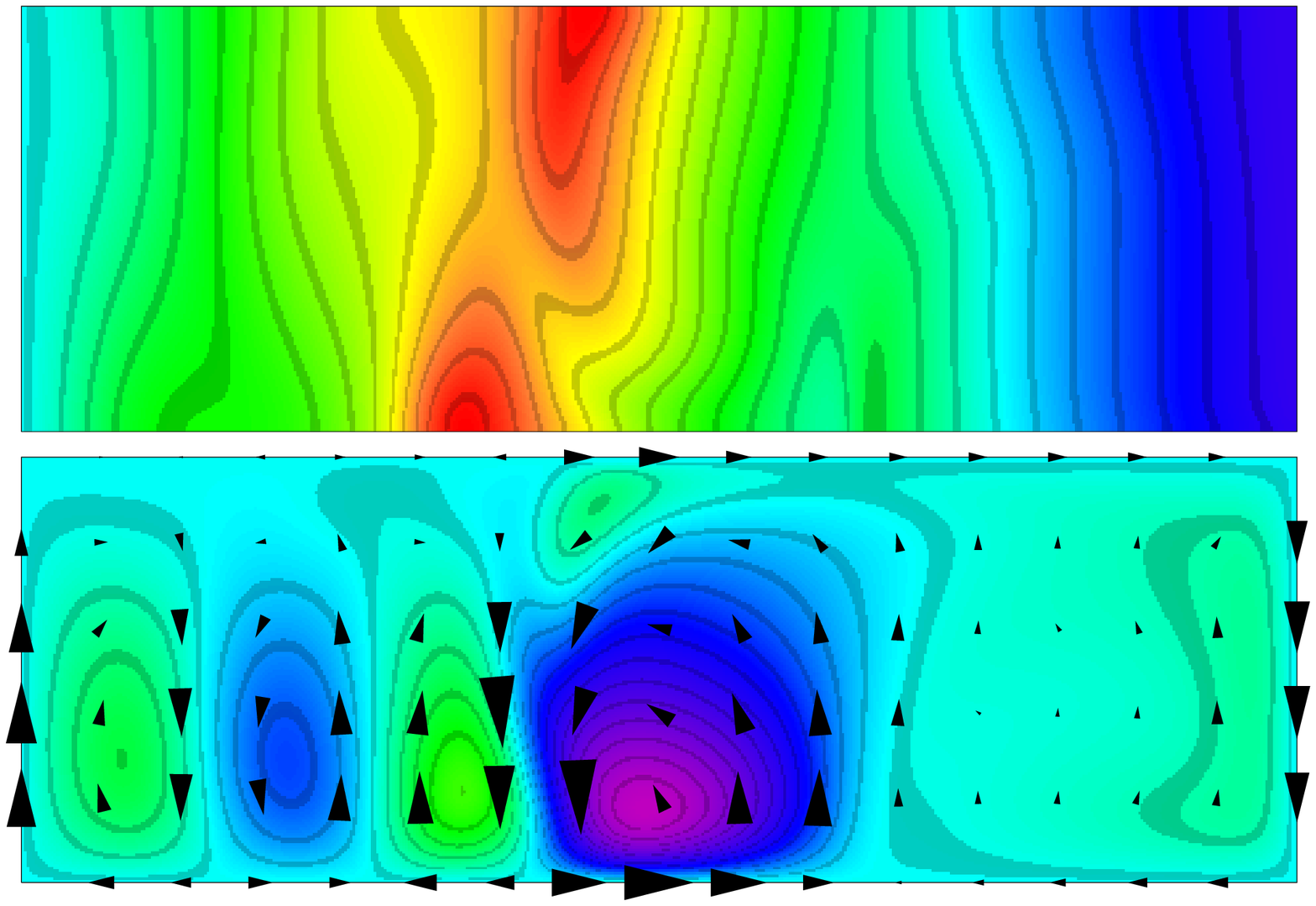}}
\end{minipage}
\end{minipage}
}
\caption{Results with $Q=128$, $\Omega=0.3$ and $\theta=20$ and a 
         constant $T$ as bottom boundary.
         Compared to the results for $\theta=10$ (Figure \ref{fig:Q128om03}), 
         the magnetic field is more radially dispersed, allowing 
         convection cells to form throughout the radial domain. 
         As one moves towards the outer boundary, the upflows between cells 
         grow stronger, accompanied by stronger heating above them. 
         The measured $\max|u_\phi|=0.7$, $\max|B_\phi|=1.3$, 
         $\max|\tilde{T}|=4.3$ and $j_\phi\in(-39,57)$.
         In the $(r,z)$ plane we have  
         $j_r\in(-4,15)$ and $j_z\in(-25,12)$.
         }
\label{fig:Q128om03theta20}
\end{figure*}

\begin{figure*}
\centerline{
\begin{minipage}{17cm}
\begin{minipage}{8cm}
\scalebox{0.4}
{\includegraphics{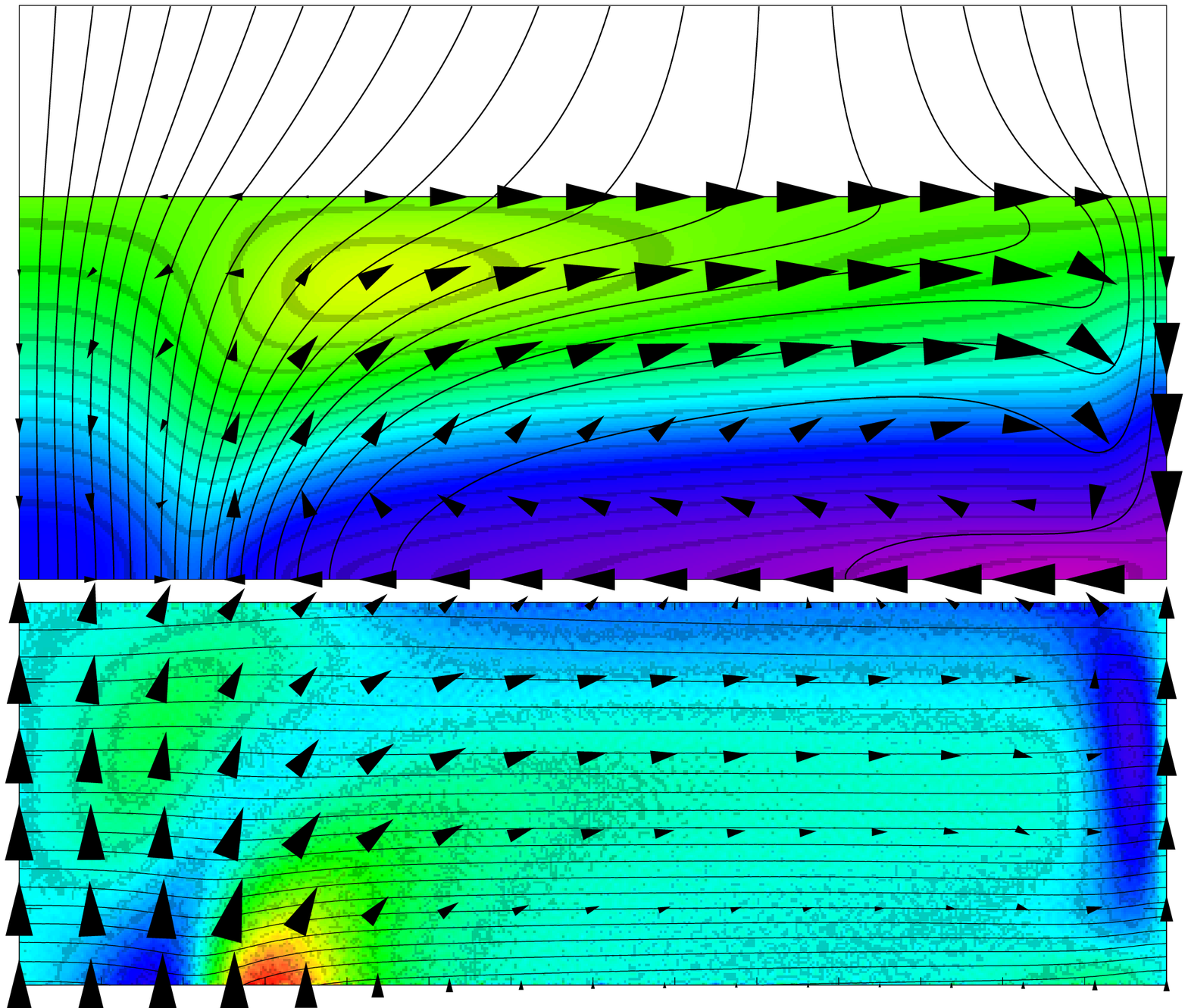}}
\end{minipage}
\begin{minipage}{8cm}
\scalebox{0.4}
{\includegraphics{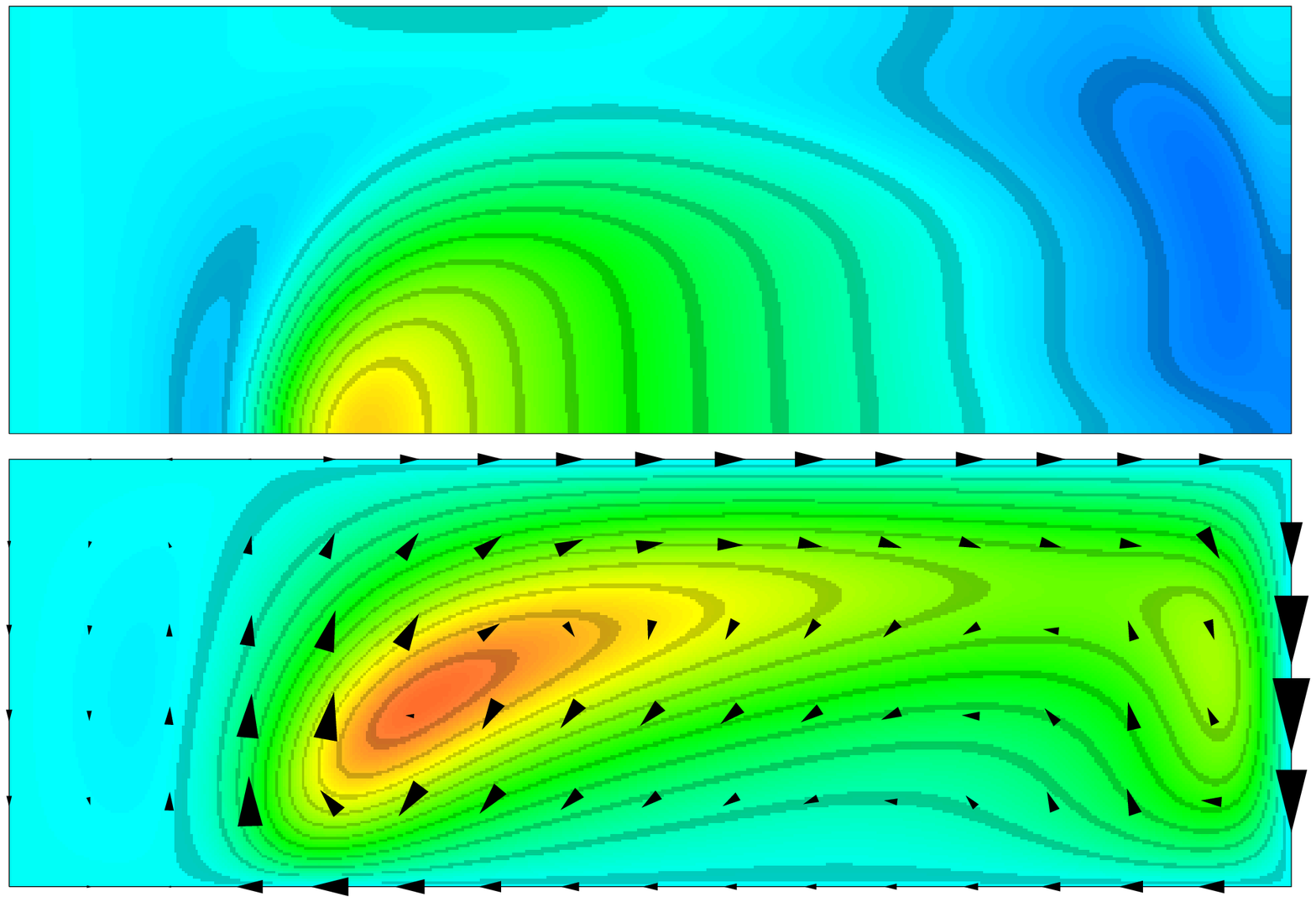}}
\end{minipage}
\end{minipage}
}
\caption{Results with $Q=32$, $\Omega=0.1$ and $\theta=20$ and 
         constant $\partial T/\partial z$ as bottom boundary.
         Compared to the results for $\theta=10$ (Figure 
         \ref{fig:Q32om01heat}), 
         the magnetic field is more radially dispersed, allowing weak 
         convection cells to form inside the flux tube. 
         The strong downflow at the outer edge cools the plasma in 
         the lower layers of the domain.  
         The measured $\max|u_r|=1.0$, $\max|u_\phi|=0.5$, 
         $\max|u_z|=1.1$, $\max|B_\phi|=2.0$, $\max|\tilde{T}|=2.2$, 
         $j_\phi\in(-41,74)$, $j_r\in(-10,7)$ and $j_z\in(-41,21)$.
         }
\label{fig:Q32om01theta20heat}
\end{figure*}

Increasing $\theta$ does not change the basic configuration of the 
numerical solution. The magnetic flux tube forming at the central axis 
remains intact, as well as the convection cells in the field-free 
region. This is true for a constant temperature as well as a constant 
$\partial T/\partial z$ bottom boundary condition. Increasing 
$\theta$ increases the strength of convection in the $(r,z)$ plane, 
as measured by the Mach number in Table \ref{tab:theta}, as well as the 
width of the magnetic flux tube. The latter can be observed in Figure 
\ref{fig:Q128om03theta20} for a constant $T$ at the bottom boundary and 
in Figure \ref{fig:Q32om01theta20heat} for $\partial T/\partial z$ constant. 
The wider flux tube leads to lower gradients in the magnetic field, 
which means the azimuthal current density around the tube, calculated 
using equation (\ref{eq:aux}), becomes weaker. All the other 
azimuthal quantities ($B_\phi$ and $u_\phi$) are also weaker when 
compared to results with $\theta=10$. 

\begin{figure}
\centerline{
\scalebox{0.46}
{\includegraphics{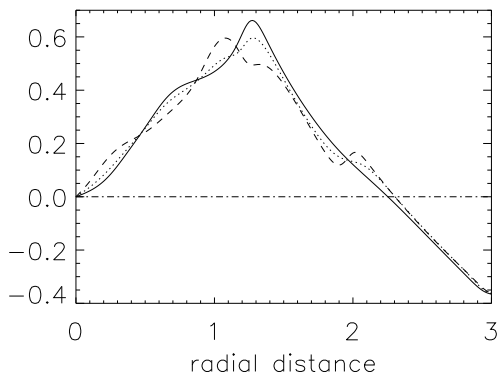}}
 }
\caption{Radial profile of $u_\phi$ corresponding to Figure 
         \ref{fig:Q128om03theta20}, with $\Omega=0.3$, $Q=128$ and 
         $\theta=20$. This should be compared with the case when 
         $\theta=10$ in Figure \ref{fig:vQ128}. 
         The lines have the same meaning as in Figure \ref{fig:vr}. 
         }
\label{fig:vQ128theta20}
\end{figure}

Figure \ref{fig:Q128om03theta20} shows a solution with $Q=128$ and 
constant $T$ at the lower boundary. When this is compared to a solution 
with the same parameter values and boundary conditions, but with 
$\theta=10$ (Figure \ref{fig:Q128om03}), one observes that the wider 
magnetic flux tube allows weak convection cells to form inside it. This 
convection is strong enough to perturb the temperature inside the flux tube 
in the top half of the numerical domain. A careful inspection of the top 
boundary shows that these convection cells cause flow along the boundary, 
so that concentric rings start to appear at the top. This is a consequence 
of the axisymmetry in our model, as one would expect cellular convection 
to form inside the flux tube. These flows of 
concentric rings around the central axis grow in size as the magnetic 
flux tube becomes wider, which occurs for higher values of $\Omega$. 
The azimuthal magnetic field has 
its maximum value in the strong collar flow next to the flux bundle. However, 
the weak convection cells inside the flux tube are defined well enough for 
$B_\phi$ to have significant components inside them (Figure 
\ref{fig:Q128om03theta20}). The direction of each convection cell determines 
the direction of the local $B_\phi$ inside it: clockwise convection contains 
a local maximum and anticlockwise a local minimum. 
The azimuthal velocity forms a Rankine vortex, as shown in Figure 
\ref{fig:vQ128theta20}. The weak convection inside the magnetic flux tube 
perturbs the rigid body rotation of the plasma inside the tube. 
The maximum value of $u_\phi$ is found next to the outer edge of the flux 
tube and a free vortex forms in the convection area. At the outer wall a 
counterflow forms, as in the case when $\theta=10$. A comparison 
between the two sets of results ( Figure \ref{fig:vQ128} for $\theta=10$ 
and Figure \ref{fig:vQ128theta20} for $\theta=20$) shows that 
max($u_\phi$) is lower and situated farther from the central axis for 
$\theta=20$. This means the counter flow at the outer boundary, generated 
because our solution has zero vertical angular momentum relative to the 
rotating reference frame, is approximately the same strength for both 
$\theta$ values. 

\begin{figure}
\centerline{
\scalebox{0.46}
{\includegraphics{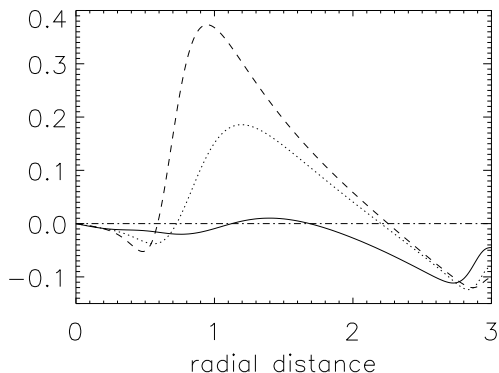}}
 }
\caption{Radial profile of $u_\phi$ corresponding to Figure 
         \ref{fig:Q32om01theta20heat}, with $\Omega=0.1$, $Q=32$ and 
         $\theta=20$. This should be compared to the case when 
         $\theta=10$ in Figure \ref{fig:vQ32heat}. 
         The lines have the same meaning as in Figure \ref{fig:vr}. 
         }
\label{fig:vQ32theta20}
\end{figure}

Figure \ref{fig:Q32om01theta20heat} shows a solution with $Q=32$ and  
$\partial T/\partial z$ constant at the lower boundary. This should 
be compared to 
Figure \ref{fig:Q32om01heat} that has the same parameter values and 
boundary conditions, but with $\theta=10$. This comparison shows that 
the magnetic flux tube is wider for $\theta=20$ and that weak convection 
occurs inside the tube. This convection is not strong enough to 
significantly heat the upper layers of the numerical domain. In fact, 
the strong downflows in the solution cools the lower layers of the 
numerical domain much more than when $\theta=10$ (Figure 
\ref{fig:heatbottom}). 
The azimuthal magnetic field is located inside the large convection 
cell next to the magnetic flux tube. As in the case for $\theta=10$, 
the maximum value of $B_\phi$ is located close to the flux tube. 
Inside the field-free convection areas we observe a free vortex, with 
its maximum next to the edge of the flux tube and a counter flow next 
to the outer wall (Figure \ref{fig:vQ32theta20}). The rotation of the 
plasma inside the flux tube is that of a rigid body, but in the opposite 
direction from the direction of the free vortex around the tube. We 
also see that the plasma inside the horizontal magnetic field on top of 
the convection zone rotates retrograde, i.e.\ in the same direction as 
the counter flow at the outside wall. This flow pattern is not a surprise, 
given the fact that we could generate retro flows at the central axis and 
the top of the numerical domain by playing with the parameter values in 
the set of results with $\theta=10$. (See the discussion in Section 
\ref{sec:Tflux}.) 

The variation in temperature is much larger for these cases with 
$\theta=20$ than in the comparable cases with $\theta=10$. For a 
constant $T$ at the bottom boundary (Figure \ref{fig:Q128om03theta20}) 
the heating occurring at the top of the numerical domain due to the strong 
upflow between the two large convection cells is three times larger than  
the heating for $\theta=10$ (Figure \ref{fig:Q128om03}). In contrast, 
the result with $\partial T/\partial z$ constant at the bottom boundary 
(Figure 
\ref{fig:Q32om01theta20heat}) shows that the strong downflow next to 
the outer wall cools the lower part of the numerical domain. The amount 
of cooling is double that which occurs for $\theta=10$ (Figure 
\ref{fig:Q32om01heat}).


\subsection{Boundary layer at the outer boundary}
\label{sec:out}
 
When comparing results with a rotating cylinder, one observes a collar 
flow next to the 
flux bundle and a clockwise convection cell at the outer boundary. 
The size of the cell at the outer boundary seems to be robust for the 
various parameter values. The only exceptions are for 
$\partial T/\partial z$ constant at the bottom boundary and with 
$\sigma\geq 0.3$ (Figure \ref{fig:Q32om01heat}), 
when the whole convection pattern change. 
Throughout the simulations we have taken care that the convection cell 
at the outer boundary does not influence the physics near the central axis 
and the flux tube.
Another effect of finite $\Omega$ is that the density contours become 
slanted at the outer boundary due to the centrifugal term in the 
Navier-Stokes equation (\ref{eq:NS}). This effect is much more noticeable 
for a constant $T$ bottom boundary condition than when 
$\partial T/\partial z$ is used. 
In contrast, the azimuthal velocity shows a sharp decrease in its 
value at the outside wall when a $\partial T/\partial z$ bottom boundary 
condition is used (Figures \ref{fig:vQ32sigma03heat} to 
\ref{fig:vQ256sigma03heat}) while the outside wall hardly register in 
the $u_\phi$ profile for a constant $T$ bottom boundary (Figure 
\ref{fig:vom03}). 
The slanted $\rho$ contours and the decrease in $u_\phi$ amplitude show  
how the influence of the outer boundary on the solution increases as 
$\Omega$ increases. Due to the formulation of the problem this is unavoidable, 
but it does not pose a problem as long as these effects stay localized at the 
outer boundary. 

In order to minimize the effect of the outer boundary on the solution, 
it was treated throughout as a slippery boundary, so that the condition 
on $u_\phi$ is given by (\ref{eq:outer}), obtained  from the off-diagonal 
elements of the rate of strain tensor (\ref{eq:strain}). 
In this paper the boundary conditions at the outer wall 
were chosen so that the coupling between the numerical domain and its 
outside surroundings is kept to a minimum. With boundary conditions 
(\ref{eq:outer}) only a vertical current exists and the Lorentz force is 
zero at the outer wall. To measure the influence of the outer wall on 
the solution, we changed its magnetic boundary condition to that of a perfect 
conductor. In this case no currents exist parallel to the wall, with a 
radial current moving through the outer wall. The condition that 
no vertical current exists leads to  
\begin{equation}
\frac{\partial B_\phi}{\partial r}=-\frac{B_\phi}{r}\: ,
\label{eq:perf}
\end{equation} 
while the radial magnetic field component stays zero, as in (\ref{eq:outer}). 
For a perfect conductor the Lorentz force has components parallel to the 
outer wall, but there is no force across the wall. This implies that 
there is a torque at the outer boundary, leading to a contribution to the 
angular momentum. 

Changing the outer boundary conditions on $B_\phi$ to (\ref{eq:perf}) 
changes the solution slightly, but only when the azimuthal amplitudes 
at the boundary become significant when compared to the solution near the 
central axis, as was the case for $\partial T/\partial z$ constant  
at the bottom boundary and $\sigma\geq 0.3$ (Figure \ref{fig:Q32om01heat}).
When the electrically insulating outer wall (Figure \ref{fig:Q32om01heat}) 
is compared with a perfectly conducting wall,  
the solution is only slightly perturbed close to the outer boundary. 
The boundary conditions of the bottom boundary, described in 
(\ref{eq:bottom}), allow currents parallel to the lower boundary 
but not through it. Ditto for the Lorentz force. As a result, in all 
simulations 
with a perfect conductor at the outer boundary, the largest current entering 
the numerical domain is situated in the bottom right hand corner.
We attempted to change the magnetic boundary condition on the lower boundary, 
but found that the solution is highly sensitive to any changes and becomes 
numerically unstable. 
Changing the boundary condition of $B_\phi$ on the outer boundary 
left the azimuthal velocity field intact.  

The difference between boundary conditions (\ref{eq:outer}) and 
(\ref{eq:perf}) was thoroughly tested. We started numerical runs from 
a uniform magnetic field for both sets of conditions, which lead to 
almost identical time independent solutions. The numerical results presented 
in this paper were started with (\ref{eq:perf}) and then continued 
with (\ref{eq:outer}). In all cases no significant difference between the 
numerical solutions could be observed. 


\begin{table*}
 \centering
 \begin{minipage}{173mm}
  \caption{Survey of numerical solutions obtained with $T$ constant at 
           the lower boundary and with parameters  
           $R=10^5$, $\zeta_0=0.2$, $m=1$, $\gamma=5/3$, $\Gamma=3$. 
           The star superscript in the $\Omega$ column indicates time 
           dependent solutions.}
  \label{tab:surveyT}
  \begin{tabular}{@{}lrlrrccccccc@{}}
  \hline
   $\Omega$ &  $Q$ & $\sigma$ & $\theta$ & $Ro$ & 
   $\max|\tilde{T}|$ &  max(Mach) & $\max|u_\phi|$ & 
   $\max|B_r|$ & $\max|B_\phi|$ & $\max|B_z|$ & $j_\phi$ range \\
  \hline
  0$^*$ & 32 & 1 & 10 & $\infty$ & 
          (2.9,3.1) & (1.2,1.3) & $\rightarrow 0$ & (13,16.7) & 
                      $\rightarrow 0$ & (30.5,38.8) & (-465,555) \\
  0$^*$ & 128 & 1 & 10 & $\infty$ & 
          (2.8,2.9) & (1.0,1.2) & $\rightarrow 0$ & (7.3,7.9) & 
                            $\rightarrow 0$ & (14.6,16.9) & (-184,205) \\
  0    & 256 & 1   & 10 & $\infty$ & 2.7 & 0.9 & $\rightarrow 0$ &  5.3 & $\rightarrow 0$ 
                                               & 10.8 & (-80,135) \\
  0.1  & 32  & 0.3 & 10 &  42.2 & 3.7 & 1.7 & 1.6 & 17.8 & 5.5 & 43.2 & (-229,445) \\
  0.1  & 32  & 0.6 & 10 &  59.8 & 3.2 & 1.1 & 1.1 & 15.2 & 4.5 & 37.0 & (-208,393) \\
  0.1  & 32  & 1   & 10 &  77.5 & 2.9 & 0.9 & 0.9 & 13.3 & 4.0 & 31.7 & (-189,365) \\
  0.1  & 128 & 0.3 & 10 &  42.2 & 3.3 & 1.7 & 0.8 & 8.6  & 1.7 & 18.9 & (-107,192) \\
  0.1  & 128 & 0.6 & 10 &  59.8 & 3.0 & 1.1 & 0.6 & 7.6  & 1.6 & 15.9 &  (-86,171) \\
  0.1  & 128 & 1   & 10 &  77.5 & 2.8 & 0.8 & 0.5 & 6.9  & 1.5 & 14.2 &  (-90,178) \\
  0.1  & 128 & 1   & 20 & 205.5 & 4.3 & 1.0 & 0.3 & 3.8  & 0.6 & 7.9  &   (-38,67) \\
  0.1  & 256 & 0.3 & 10 &  42.2 & 2.6 & 1.1 & 0.6 & 5.2  & 0.6 & 11.2 &  (-65,103) \\
  0.1  & 256 & 0.6 & 10 &  59.8 & 2.8 & 1.0 & 0.4 & 5.0  & 1.0 &  9.5 &  (-69,112) \\
  0.1  & 256 & 1   & 10 &  77.5 & 2.7 & 0.7 & 0.3 & 4.7  & 0.9 &  8.9 &  (-61,107) \\
  0.1$^*$  & 256 & 1   & 20 & 205.8 & 
         (3.8,3.9) & 0.6 & (0.1,1.5) & (2.0,2.0)  & 0.2  & (7.0,7.1)  &   (-33,39) \\
  0.2  & 32  & 1   & 10 &  38.7 & 2.9 & 0.9 & 1.3 & 10.7 & 4.2 & 24.4 & (-148,271) \\
  0.2  & 32  & 1   & 20 & 102.8 & 4.7 & 1.2 & 1.3 & 7.4  & 2.7 & 17.7 &  (-74,153) \\
  0.2  & 128 & 1   & 10 &  38.7 & 2.8 & 0.8 & 0.8 & 6.3  & 2.6 & 12.8 &  (-73,149) \\
  0.2  & 128 & 1   & 20 & 102.8 & 4.3 & 1.0 & 0.6 & 3.7  & 1.1 & 7.5  &   (-37,64) \\
  0.2  & 256 & 1   & 10 &  38.7 & 2.7 & 0.7 & 0.6 & 4.4  & 1.7 & 8.4  &   (-56,101) \\
  0.2$^*$  & 256 & 1   & 20 & 102.8 & 
          (3.7,3.9) & (0.6,0.7) & 0.3 & (1.8,2.0)  & (0.3,0.4) & (4.5,4.6)  & (-32,37) \\
  0.3  & 32  & 1   & 10 &  25.8 & 2.9 & 0.9 & 1.5 & 8.5  & 8.0 & 20.3 & (-143,220) \\
  0.3  & 32  & 1   & 20 &  68.5 & 4.6 & 1.2 & 1.5 & 6.3  & 2.8 & 14.3 &  (-66,123) \\
  0.3  & 128 & 1   & 10 &  25.8 & 2.7 & 0.8 & 1.0 & 5.3  & 2.9 & 10.6 & (-65,125) \\
  0.3  & 128 & 1   & 20 & 68.5 & 4.3 & 1.0 & 0.7 & 3.3  & 1.3 & 7.8  & (-39,57) \\
  0.3  & 256 & 1   & 10 & 25.8 & 2.6 & 0.6 & 0.7 & 3.8  & 2.1 & 7.2  & (-55,89) \\
  0.3$^*$ & 256 & 1 & 20 & 68.5 & 
         3.7 & 0.5 & (1.6,1.7) & (1.6,1.7) & 0.4 & 3.8 & (-31,37) \\
  \hline
  \end{tabular}
 \end{minipage}
\end{table*}

\begin{table*}
 \centering
 \begin{minipage}{173mm}
  \caption{Survey of numerical solutions obtained with $\partial T/\partial z$ constant 
           at the lower boundary and with parameters  
           $R=10^5$, $\zeta_0=0.2$, $m=1$, $\gamma=5/3$, $\Gamma=3$.
           The star superscript in the $\Omega$ column indicates time dependent solutions.}
  \label{tab:surveyFlux}
  \begin{tabular}{@{}lrlrrccccccccc@{}}
  \hline
   $\Omega$ &  $Q$ & $\sigma$ & $\theta$ & $Ro$ & 
   $\max|\tilde{T}|$ &  max(Mach) & $\max|u_\phi|$ & 
   $\max|B_r|$ & $\max|B_\phi|$ & $\max|B_z|$ & $j_\phi$ range\\
  \hline
  0 & 32  & 1 & 10 & $\infty$ & 1.2 & 0.6 & $\rightarrow 0$ & 6.3 & $\rightarrow 0$
                                          & 21.8 & (-73,142) \\
  0 & 128 & 1 & 10 & $\infty$ & 1.1 & 0.4 & $\rightarrow 0$ & 4.3 & $\rightarrow 0$ 
                                          & 13.4 & (-46,92) \\
  0 & 256 & 1 & 10 & $\infty$ & 1.1 & 0.4 & $\rightarrow 0$ & 3.7 & $\rightarrow 0$ 
                                          & 9.8  & (-38,72) \\
  0.1 & 32  & 0.1 & 10 &  23.7 & 1.4 & 1.3 & 0.8 & 6.3 & 3.2 & 16.9 & (-73,105) \\
  0.1 & 32  & 0.3 & 10 &  42.2 & 1.2 & 1.1 & 0.5 & 4.3 & 3.0 & 17.1 & (-56,106) \\
  0.1 & 32  & 0.6 & 10 &  59.8 & 1.1 & 0.6 & 0.4 & 5.0 & 3.8 & 31.0 & (-56,111) \\
  0.1 & 32  & 1   & 10 &  77.5 & 1.1 & 0.6 & 0.5 & 5.8 & 3.7 & 19.7 & (-59,149) \\
  0.1 & 32  & 1   & 20 & 205.5 & 2.2 & 0.8 & 0.5 & 4.3 & 2.0 & 14.3 &  (-41,74) \\
  0.1 & 128 & 0.1 & 10 &  23.7 & 1.4 & 1.5 & 0.5 & 3.9 & 1.5 & 12.5 &  (-38,80) \\
  0.1 & 128 & 0.3 & 10 &  42.2 & 1.2 & 0.8 & 0.4 & 3.8 & 1.5 & 11.2 &  (-46,78) \\
  0.1 & 128 & 0.6 & 10 &  59.8 & 1.1 & 0.6 & 0.4 & 4.1 & 1.6 & 11.4 &  (-48,87) \\
  0.1 & 128 & 1   & 10 &  77.5 & 1.1 & 0.5 & 0.4 & 4.2 & 1.6 & 11.8 &  (-49,106) \\
  0.1 & 128 & 1   & 20 & 205.5 & 1.9 & 0.5 & 0.3 & 2.6 & 0.8 & 7.2  &  (-25,40) \\
  0.1 & 256 & 0.1 & 10 &  23.7 & 1.3 & 1.1 & 0.4 & 3.3 & 1.0 &  9.3 &  (-34,65) \\
  0.1 & 256 & 0.3 & 10 &  42.2 & 1.1 & 0.6 & 0.4 & 3.2 & 1.0 &  8.3 &  (-37,63) \\
  0.1 & 256 & 0.6 & 10 &  59.8 & 1.1 & 0.5 & 0.4 & 3.2 & 1.1 &  8.3 &  (-37,59) \\
  0.1 & 256 & 1   & 10 &  77.5 & 1.1 & 0.4 & 0.3 & 3.3 & 1.1 &  8.4 &  (-37,63) \\
  0.1 & 256 & 1   & 20 & 205.5 & 1.6 & 0.2 & 0.1 & 1.5 & 0.4 &  4.8 &  (-15,29) \\
  0.2 & 32  & 1   & 10 &  38.7 & 1.2 & 0.5 & 0.8 & 3.3 & 4.4 & 21.5 &  (-56,95) \\
  0.2 & 32  & 1   & 20 & 102.8 & 2.2 & 0.7 & 0.7 & 2.9 & 3.0 & 9.9  &  (-44,54) \\
  0.2 & 128 & 1   & 10 & 38.7  & 1.3 & 0.4 & 0.6 & 2.8 & 2.4 & 12.2 &  (-43,67) \\
  0.2 & 128 & 1   & 20 & 102.8 & 1.9 & 0.7 & 0.5 & 2.0 & 1.4 & 6.5  & (-24,35) \\
  0.2$^*$ & 256 & 1 & 10 & 38.7 & 
         (1.1,1.5) & 0.3 & 0.5 & 2.2 & (1.8,1.9) & (5.6,12.8) & (-35,57) \\
  0.2 & 256 & 1   & 20 & 102.8 & 1.6 & 0.2 & 0.3 & 1.4 & 0.7 & 4.3  & (-15,26) \\
  0.3 & 32  & 1   & 10 &  25.8 & 1.2 & 0.5 & 0.9 & 2.6 & 4.8 & 24.9 & (-54,110) \\
  0.3 & 32  & 1   & 20 &  68.5 & 2.1 & 0.6 & 0.9 & 2.2 & 3.4 & 10.2 & (-37,45) \\
  0.3$^*$ & 128 & 1   & 10 &  25.8 & 
         (1.1,1.3) & (0.3,0.4) & 0.7 & (1.8,2.3) & (2.6,2.8) & (6.6,14.2) & (-40,62) \\
  0.3 & 128 & 1   & 20 &  68.5 & 1.9 & 0.4 & 0.6 & 1.5 & 1.8 & 7.0  & (-24,27) \\
  0.3$^*$ & 256 & 1   & 10 &  25.8 & 
         1.1 & (0.2,0.3) & 0.5 & (1.4,1.8) & (1.9,2.0) & (6.8,7.4) & (-33,46) \\
  0.3 & 256 & 1   & 20 &  68.5 & 1.6 & 0.2 & 0.3 & 1.1 & 0.9 & 4.1  & (-18,21) \\
  \hline
  \end{tabular}
 \end{minipage}
\end{table*}


\subsection{Time dependence}
\label{sec:time}

Increasing the angular velocity $\Omega$ increases the width of the 
magnetic flux tube and allows weak convection cells to form inside the 
tube, similar to those formed in Figure \ref{fig:Q128om03theta20}.    
If the value of $\Omega$ becomes large enough, these cells undergo 
periodic motion. For a constant $T$ bottom boundary 
condition, $Q=256$ and $\theta=20$, the weak convection cells inside the 
flux tube oscillate in the radial direction. As $\Omega$ increases from 
0.1 to 0.3, the amplitude of this oscillation increases as well.  
In contrast, for a bottom boundary condition of constant 
$\partial T/\partial z$ with $\Omega=0.3$ and $\theta=10$, a hot blob 
forms due to weak upflows inside the tube. This blob then moves towards 
the central axis where it dissipates. A new blob forms inside the flux 
tube and the process repeats itself. This happens for $Q$ values of 
128 and 256 with $\Omega=0.3$, as well as for $Q=256$ and $\Omega=0.2$. 
When the value of $\theta$ is doubled to 20, the solutions become time 
independent again. 

When the forced conservation of $L_z$ is lifted, its value tends to 
drift, as discussed in Section \ref{sec:num}. In the case of a time 
dependent solution, $L_z$ oscillates in sympathy with the oscillation 
in the convection, in addition to its drift. The amplitude of the 
oscillation can thus be expressed in terms of an equivalent solid body 
rotation, which is of order $O(10^{-2})$. This compares to a drift in 
the value of $L_z$ of order $O(10^{-5})$ per unit time. 


\section{Summary}

We have investigated magnetoconvection around a magnetic flux bundle 
in a cylinder, when the cylinder is rotated at a constant angular 
velocity $\Omega$. The model uses a compressible plasma with 
density and temperature gradients simulating the upper solar convection 
zone. All the numerical solutions that we obtained are presented in Tables 
\ref{tab:surveyT} and \ref{tab:surveyFlux}. Throughout the calculations the 
maximum velocities are in the $(r,z)$ plane, so the the maximum Mach number 
in these tables are a good proxy for $u_r$ and $u_z$. For time dependent 
solutions we present the range in which the different diagnostics lie.

With no rotation ($\Omega=0$) and a constant temperature at the lower 
boundary, the solution is in the form of a flux tube situated at 
the central axis, surrounded by a field-free annular convection ring that 
forms a collar around the flux tube (Section \ref{sec:norot}). 
This magnetic configuration lends itself to the description of idealized 
pores and sunspots. The collar flow has been measured in the convection 
around both phenomena. (See \citet{BothaEA06} and references in it.)  

The introduction of a constant angular velocity $\Omega$ widens the 
magnetic flux tube (Section \ref{sec:rot1}). Other ways to increase 
the tube width are to increase the magnetic field strength (Section 
\ref{sec:mag}) and to increase the heat flux into the numerical domain 
from below (Section \ref{sec:Tflux}). 
If the magnetic field strength (i.e.\ $Q$) is kept constant and the 
tube width is increased by means of one of the above, then the 
amplitude of the vertical magnetic field component in the flux tube 
is lowered. This allows weak convection cells to form inside the tube. As 
$\Omega$ increases, the flux tube widens and the weak convection becomes 
stronger so that eventually concentric rings appear at the top of the 
numerical domain (Figures \ref{fig:Q32om03} and \ref{fig:Q128om03theta20}).
In a fully 3D model one would expect cellular convection to form inside 
the flux tube.  

Increasing $\Omega$ also brings time dependence to the solution 
(Section \ref{sec:time}). For moderate $\Omega$ values the weak 
convection cells oscillate horizontally inside the magnetic flux tube, 
while for large $\Omega$ values the weak cells push periodically through 
the edge of the flux tube into the field-free convection area. 
This time dependence can be reduced by increasing the strength of the 
magnetic field (Section \ref{sec:mag}).

The collar flow around the magnetic flux tube is influenced by the strength 
of the convection and the temperature prescription at the lower boundary. 
By lowering the value of the Prandtl number ($\sigma$), the convection 
becomes stronger and the size of the collar cell increases 
(Section \ref{sec:Prandtl}). The stronger convection pushes the magnetic 
flux tighter at the central axis so that the flux tube width decreases and  
the magnetic field strength on axis increases. For $\sigma=0.1$ the collar 
flow survives a change of the lower boundary condition from 
a constant temperature to a constant $\partial T/\partial z$ 
(Section \ref{sec:Tflux}). However, for $\sigma\geq 0.3$ the collar flow 
is destroyed and the magnetic field is dragged away from the central axis 
(Figure \ref{fig:Q32om01heat}). 
Weak convection cells form inside this wider flux tube. 

The azimuthal velocity and magnetic fields are driven by the imposed 
$\Omega$, because in the absence of rotation these quantities have 
very small amplitudes, generated by the initial plasma perturbation, 
which decay exponentially to zero with time (Section \ref{sec:norot}). 
It follows that as $\Omega$ increases, the magnitudes of $u_\phi$ and 
$B_\phi$ increase (Figure \ref{fig:summary}). In contrast, the 
amplitudes of $u_r$ and $u_z$ hardly change with $\Omega$.  
For all values of $\Omega$ the azimuthal flow pattern fits that of a 
Rankine vortex: 
in areas with strong magnetic field the plasma tends to rotate as a rigid 
body while around it a free vortex forms in the field-free convection areas. 
This means that $\max(u_\phi)$ is located outside the flux tube edge. 
A finite $\Omega$ shortens the wavelength of convection in the 
radial direction, so that the initial convection annulus breaks up into 
more than one convection cell (Section \ref{sec:rot1}). The vortex forming 
around the flux tube is not dependent on the number of convection cells in 
the field-free region. 

The plasma inside the magnetic flux tube and the vortex around the 
tube flow prograde relative to the rotating cylindrical reference frame 
(Figure \ref{fig:vphi}). A retrograde or counter flow appears next to the 
outer wall of the cylinder. This counter flow is due to the fact that 
in our solution the vertical component of the angular momentum is zero 
relative to the rotating reference frame.  
We initialize the simulations with $L_z=0$ and the counter 
flow appears at the outer wall to maintain the status quo. 
To obtain a retrograde flow at the central axis like 
\citet{JonesGalloway93}, we have to change the bottom boundary condition 
on the temperature from constant $T$ to constant $\partial T/\partial z$ 
(Section \ref{sec:Tflux}). This change in boundary condition 
also creates a strong horizontal magnetic component in the top layers 
of the numerical domain, which may rotate retrograde with the plasma 
at the axis and the outer wall (Figure \ref{fig:vQ256sigma03heat}). 
Alternatively, 
by generating weak turbulence inside the magnetic flux tube, 
it is possible to perturb the rigid body rotation of the plasma inside 
the flux tube to such an extent that one gets prograde and retrograde flow 
inside the flux tube. This is more likely to happen with a constant 
$\partial T/\partial z$ than for a constant $T$ lower boundary condition. 

Unlike the azimuthal velocity, the azimuthal magnetic field is influenced 
by the structure of the convection cells. Max($B_\phi$) is confined to 
the strongest convection cell closest to the outer edge of the magnetic 
flux tube. For a constant $T$ lower boundary condition this is usually 
the collar flow next to the magnetic flux tube. For constant 
$\partial T/\partial z$ as lower boundary condition and $\sigma=0.1$, 
significant parts of $B_\phi$ form inside the weak convection in the flux 
tube as well as inside collar cell outside the tube (Figure 
\ref{fig:Q32om01sigma01heat}). For constant $\partial T/\partial z$ and 
$\sigma\geq 0.3$, the $B_\phi$ forms in the large convection cell 
around the flux tube, with a local maximum next to the flux tube 
(Figure \ref{fig:Q32om01heat}).
The direction of $B_\phi$ depends on the convection direction; 
for anticlockwise flow (as in the collar flow around the flux tube) it 
points in the negative $\phi$ direction and vice versa for clockwise flow. 
When the solution has one large convection cell with clockwise flow, 
which we obtain with a constant $\partial T/\partial z$ at the bottom 
boundary, a local $\max|B_\phi|$ forms at the outer wall, but its radial 
width and amplitude is small so that it does not influence the numerical 
solution inside the domain (Section \ref{sec:Tflux}).

The current density in the $(r,z)$ plane always forms around the local 
$\max|B_\phi|$ and flows in the same direction as the local convection. 
The azimuthal current density forms around the edge of the flux tube where 
the magnetic field lines have the largest gradients and curvature. 
This means any process 
that widens the flux tube, i.e.\ straightens the magnetic field lines, 
will decrease $j_\phi$ and vice versa. Increasing $\Omega$ 
(Section \ref{sec:rot2}), the magnetic field strength (Section  
\ref{sec:mag}), the stratification in the domain (Section 
\ref{sec:depth}), and changing the temperature lower boundary 
condition to constant $\partial T/\partial z$ (Section \ref{sec:Tflux}) 
lead to a decrease in the amplitude of $j_\phi$, while a lower Prandtl number 
(Figure \ref{fig:Q32om01sigma03}) increases the amplitude of $j_\phi$. 
When the weak convection inside the magnetic flux tube becomes strong 
enough to bend the field lines, local maxima of $|j_\phi|$ starts to form. 

Lowering the Prandtl number ($\sigma$) increases the strength of 
convection (Section \ref{sec:Prandtl}) as well as thermal diffusivity. 
Thus stronger upflows lead to stronger localized heating in the upper layer, 
while the radial extent of the heated plasma is reduced 
(Figure \ref{fig:Q32om01sigma03}). The stronger convection also causes 
significant variations in the density gradient inside the field-free 
convection area. In contrast, a finite $\Omega$ with $\sigma=1$ has 
little effect on the density inside the convection area (Section 
\ref{sec:rot2}). Only at the outer 
boundary does the rotation change the density gradient in a significant 
way, but the radial extent of this layer is small and does not influence 
the rest of the domain. Inside the magnetic flux tube the density is 
relatively unaffected for $\Omega\leq 0.3$. Relative large $\Omega$ values 
are necessary to observe a significant influence by the centrifugal force. 

To ascertain the effect of the lower boundary, 
we changed the temperature boundary condition (Section \ref{sec:Tflux}) 
and the stratification in the numerical domain (Section \ref{sec:depth}). 
Increasing the stratification effectively increases the heat flux 
through the lower boundary into the domain. This widens the magnetic flux 
tube, allowing weak convection cells to form inside it. 
However, the convection in the field-free regions and the configuration 
of the magnetic field stay essentially the same (Table \ref{tab:theta} 
and Figures \ref{fig:Q128om03theta20} and \ref{fig:Q32om01theta20heat}). 
In contrast, changing the temperature prescription from a constant 
temperature to constant $\partial T/\partial z$ drastically affected the 
solution. 
The bottom temperature reduces slightly (Figure \ref{fig:heatbottom}), 
but this does not account for the changes in the solution. The amplitude 
of the convection reduces significantly (Table \ref{tab:T}) and for 
$\sigma\geq 0.3$ the flow pattern and magnetic field configuration 
change radically (Figures \ref{fig:Q32om01sigma01heat} and 
\ref{fig:Q32om01heat}).


\begin{figure}
\centerline{
\scalebox{0.7}
{\includegraphics{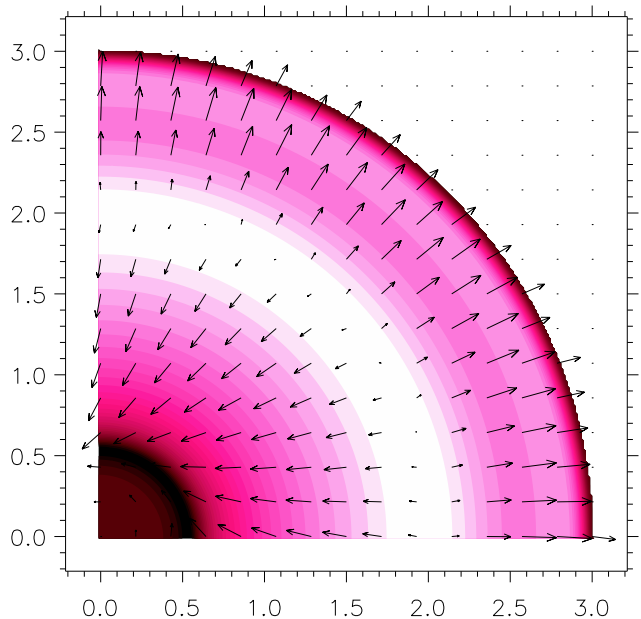}}
 }
\caption{Velocity field on the ($r,\phi$) plane at $z=0.25$ for $Q=32$, 
         $\sigma=1$, $\Omega=0.1$, and constant $T$ bottom boundary, 
         shown in Figure \ref{fig:Q32om01}. 
         Arrows represent $u_r$ and $u_\phi$ and colour the sound speed 
         perturbation $\tilde{c}_s$.  
         Max($\tilde{c}_s$)=0.54 is the light shade between the two 
         convection cells at $r=1.9$ and min($\tilde{c}_s$)=-0.03 the dark 
         shade on the edge of the magnetic flux tube at $r=0.5$.  
         }
\label{fig:spinT}
\end{figure}

\begin{figure}
\centerline{
\scalebox{0.7}
{\includegraphics{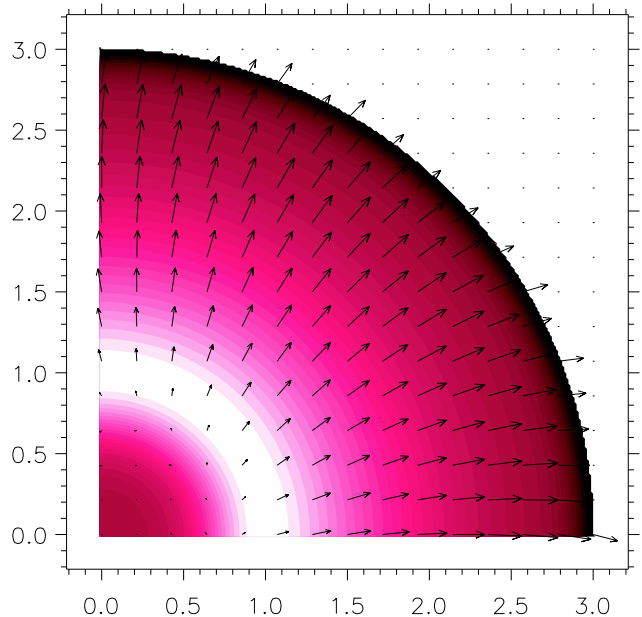}}
 }
\caption{Velocity field on the ($r,\phi$) plane at $z=0.25$ for $Q=32$, 
         $\sigma=1$, $\Omega=0.1$, and constant $\partial T/\partial z$ 
         bottom boundary, shown in Figure \ref{fig:Q32om01heat}. 
         The diagnostics has the same meaning as in Figure \ref{fig:spinT}.
         Max($\tilde{c}_s$)=0.04 is the light shade around the magnetic flux 
         tube at $r=1$ and min($\tilde{c}_s$)=-0.12 the dark shade at the 
         outer wall at $r=3$.
         }
\label{fig:spindTdz}
\end{figure}


The numerical solutions obtained in this study point to a specific radial 
profile for azimuthal velocities in sunspots that rotate around their 
own axis. Inside the umbra, where the vertical magnetic field component is 
strong, the plasma rotates as a rigid body while the convection around the 
umbra is in the form of a vortex. This profile is supported by most of the 
observations. Photospheric observations place the maximum azimuthal velocity 
inside the penumbra, while helioseismic observations show a vortex flowing 
around the flux tube in the convection zone. The typical azimuthal velocity 
$(u_\phi)$ in the photosphere is of the order $10^{-2}\;\mbox{km s}^{-1}$ 
\citep{BrownEA03}. \citet{ZhaoKosovichev03} measured 
$\max(u_\phi)\approx 0.5\;\mbox{km s}^{-1}$ below the photosphere at depths 
0-3 and 9-12 Mm. This compares well with our measured velocities in  
Tables \ref{tab:surveyT} and \ref{tab:surveyFlux}, where for low 
angular velocity ($\Omega=0.1$) we obtain 
$\max(u_\phi)\approx O(10^{-1})\;\mbox{km s}^{-1}$, 
taking a sound speed of 1.29 $\mbox{km s}^{-1}$ as reference speed. 

We present Figures \ref{fig:spinT} and \ref{fig:spindTdz} to facilitate 
comparison of our results with local helioseismic measurements, of 
which Figures 6 to 8 in \citet{Kosovichev02} are examples. Figure 
\ref{fig:spinT} shows the flow in the $(r,\phi)$ plane for a constant 
$T$ and Figure \ref{fig:spindTdz} for a constant $\partial T/\partial z$ 
bottom boundary. These planes correspond to depth $z=0.25$ in Figures 
\ref{fig:Q32om01} and \ref{fig:Q32om01heat} respectively, so that 
Figure \ref{fig:spinT} represents two convection cells with a collar flow 
around a well-defined magnetic flux tube, while Figure \ref{fig:spindTdz} 
represents one outflowing convection cell that drags the magnetic field 
lines away from the central axis. 
The arrows show that $u_r$ dominates $u_\phi$ for $\Omega=0.1$, with 
azimuthal flow patterns more visible in the inner radius closer to the 
magnetic flux tube. The size of $u_\phi$ relative to $u_r$ increases with 
$\Omega$ (Figure \ref{fig:summary}), so that the Rankine vortex becomes more 
visible for higher values of $\Omega$. In Figures \ref{fig:spinT} and 
\ref{fig:spindTdz} the outer boundary condition $u_r=0$ still holds, with 
arrows chosen close to the boundary showing that the flow has finite 
size next to the boundary. 
The colour palette shows the perturbed sound speed ($c_s$)in the plane.
Where there is an upwelling the plasma is heated and vice versa. Figure 
\ref{fig:spinT} shows the warmer plasma -- and hence higher sound speed -- 
between the two convection cells and Figure \ref{fig:spindTdz} at the 
upflow next to the magnetic flux tube. Downflow with its accompanied cooler 
plasma -- and hence lower sound speed -- occurs around the edge of the 
flux tube for Figure \ref{fig:spinT} and at the outer wall for Figure 
\ref{fig:spindTdz}. The flux tube itself is cooler than the rest of the 
surrounding plasma (Figure \ref{fig:spinT}), but where weak convection 
inside it exists, it starts to heat up (Figure \ref{fig:spindTdz}). 
The difference in max($c_s$) between Figures \ref{fig:spinT} and 
\ref{fig:spindTdz} is thus due to the radical different flow pattern for 
each case. 

We generate azimuthal flow in this axisymmetric model by rotating the 
cylinder around its axis at a constant angular velocity. As a result we 
obtain time independent solutions, in contrast to the highly time dependent 
observations. For low angular velocities the flow inside the magnetic flux 
tube and the vortex flow are prograde. Due to the fact that our model 
conserves the vertical component of the angular momentum, a retrograde 
flow appears next to the outer wall. We  
find that high angular velocities tend to break the umbra into 
concentric rings and introduce time dependence in the form of periodic 
behaviour in the radial direction. These phenomena have not been 
observed in sunspots and may be due to the axisymmetry in our model. 
It is more likely that our numerical results obtained with low angular 
velocities are realistic models of solar observations. 


\section{Acknowledgments}

We would like to thank Chris Jones for informative discussions. 
GJJB and FHB would like to acknowledge financial support from NASA grant 
NNG 04GG07G. GJJB would also like to acknowledge support through 
PPARC grant PPA/G/O/2002/00014 and STFC grant PP/E001092/1. 
NEH would like to acknowledge support from NASA grants NNG06GD45G and 
NNM07AA01C.


\label{lastpage}

\end{document}